\documentclass[twocolumn,hyphens,spaces,obeyspaces]{aastex63}
\usepackage{xcolor}
\usepackage{courier}
\usepackage[T1]{fontenc}
\usepackage{mathtools}
\usepackage{longtable}
\usepackage{xspace}
\usepackage{gensymb}
\usepackage[para]{threeparttable}
\usepackage{url}

\title{Swift-XRT and NuSTAR Monitoring of Obscuration Variability in Mrk 477}

\shorttitle{Obscuration Variability monitoring of Mrk 477}
\shortauthors{Torres-Alb\`a et al.}

\newcommand{\swift}{\textit{Swift}-XRT\xspace}
\newcommand{\myt}{\texttt{MYTorus}\xspace}
\newcommand{\bor}{\texttt{borus02}\xspace}
\newcommand{\uxc}{\texttt{UXCLUMPY}\xspace}
\newcommand{\xmm}{\textit{XMM-Newton}\xspace}
\newcommand{\nustar}{\textit{NuSTAR}\xspace}
\newcommand{\chandra}{\textit{Chandra}\xspace}

\newcommand{\cf}{C$_{\rm f}$\xspace}
\newcommand{\ctk}{C$_{\rm TK}$\xspace}
\newcommand{\nhav}{N$_{\rm H,av}$\xspace}
\newcommand{\nhlos}{N$_{\rm H,los}$\xspace}
\newcommand{\sigtor}{$\sigma_{\rm tor}$\xspace}
\newcommand{\iang}{$\theta_{\rm obs}$\xspace}
\newcommand{\fscat}{$F_{\rm s}$\xspace}
\newcommand{\cnorm}{$C_{\rm norm}$\xspace}

\begin{document}
\title{Swift-XRT and NuSTAR Monitoring of Obscuration Variability in Mrk477}

\author[0000-0003-3638-8943]{N. Torres-Alb\`{a}}
\altaffiliation{GECO Fellow}
\affiliation{Department of Physics and Astronomy, Clemson University, Kinard Lab of Physics, Clemson, SC 29634, USA}
\affiliation{Department of Astronomy, University of Virginia, P.O. Box 400325, Charlottesville, VA 22904, USA}

\author[0009-0009-3413-5919]{Z. Hu}
\affiliation{Department of Physics, University of Miami, Coral Gables, FL 33124, USA}

\author[0000-0003-2287-0325]{I. Cox}
\affiliation{Department of Physics and Astronomy, Clemson University,  Kinard Lab of Physics, Clemson, SC 29634, USA}

\author[0000-0001-5544-0749]{S. Marchesi}
\affiliation{Dipartimento di Fisica e Astronomia (DIFA), Universit\'a di Bologna, via Gobetti 93/2, I-40129 Bologna, Italy}
\affiliation{Department of Physics and Astronomy, Clemson University,  Kinard Lab of Physics, Clemson, SC 29634, USA}
\affiliation{INAF - Osservatorio di Astrofisica e Scienza dello Spazio di Bologna, Via Piero Gobetti, 93/3, 40129, Bologna, Italy}

\author[0000-0002-6584-1703]{M. Ajello}
\affiliation{Department of Physics and Astronomy, Clemson University,  Kinard Lab of Physics, Clemson, SC 29634, USA}

\author[0000-0001-6412-2312]{A. Pizzetti}
\affiliation{Department of Physics and Astronomy, Clemson University, Kinard Lab of Physics, Clemson, SC 29634, USA}
\affiliation{European Southern Observatory, Alonso de Córdova 3107, Casilla 19, Santiago 19001, Chile.}

\author[0000-0002-7825-1526]{I. Pal}
\affiliation{Department of Physics and Astronomy, Clemson University,  Kinard Lab of Physics, Clemson, SC 29634, USA}

\author[0000-0001-6564-0517]{R. Silver}
\affiliation{NASA Goddard Space Flight Center, Greenbelt, MD, 20771, USA}

\author[0000-0002-7791-3671]{X. Zhao}
\affiliation{Department of Astronomy, University of Illinois at Urbana-Champaign, Urbana, IL 61801, USA}

\begin{abstract}

We present the analysis of 15 X-ray observations of Mrk 477, a nearby Seyfert 2 active galactic nucleus, with the objective to monitor its obscuring column density variability. The full dataset consists of five archival observations, split into two \xmm, two \nustar and one \chandra observation, plus two dedicated monitoring campaigns. The monitoring campaigns were performed with \swift and \nustar, containing five observations each. We performed a simultaneous analysis using self-consistent torus models, deriving geometric properties of the torus as well as the obscuration along the line of sight. Mrk 477 is best modeled with a torus with large covering factor yet low column density (on average). Its line of sight column density oscillates between $1.5-7\times10^{23}$~cm$^{-2}$. Mrk~477 presents frequent obscuring column density variability, on timescales as short as $\sim2$~weeks. The probability of drawing a pair of obscuration-variable observations for Mrk~477 when having 2, 3, and 4 observations is 40\%, 78\% and 95\%, respectively. 
Adding the results of this work to those of another 26 sources, we find a trend of increasing obscuration variability with time (from $\sim20$\% at $\Delta t<10$~days, to $\sim60-70$\% at timescales larger than 5 years). We discuss whether this is compatible with the majority of obscuration variability coming from broad line region clouds.
\end{abstract}


\section{Introduction}\label{intro}

Active Galactic Nuclei (AGN) are supermassive black holes (SMBHs) which are actively accreting the material that surrounds them. As per the unification theory, all AGN categories are essentially distinguished by only three factors: orientation angle, intrinsic power, and the presence (or lack) of a jet \citep{Urry1995}. A key element in this categorization is the torus; a toroidal distribution of material, initially modeled as homogeneous, that surrounds the accreting SMBH, obscuring certain lines of sight. AGN viewed at an edge-on orientation are named Type-II AGN, and are generally obscured in X-rays and devoid of broad emission lines, which originate in the broad-line region (BLR). 

Later studies, both from an infrared and an X-ray perspective, paint a slightly more complex picture when it comes to the distribution of material in the torus. Spectral energy distribution (SED) fitting in the infrared favours a clumpy material distribution over a homogeneous one \citep[e.g.][]{Nenkova2002,Ramos-Almeida2014}. Similarly, changes in X-ray obscuration (measured by the obscuring line of sight, l.o.s., column density, \nhlos) in nearby AGN support an inhomogeneous torus scenario \citep[e.g.][]{Risaliti2002}. \nhlos variabiliy has been observed in timescales as short as $<1$~day \citep[e.g.][]{Elvis2004,Risaliti2009}, and as longs as years \citep[e.g.][]{Markowitz2014}. They also span a broad range of \nhlos, from $\sim10^{22}$~cm$^{-2}$ \citep[e.g.][]{Laha2020}, to Compton-thin/-thick transitions \citep[i.e. changing-look AGN, e.g.][]{Risaliti2005,Bianchi2009,Rivers2015}.

Obscuration variability in X-rays is a powerful tool, which allows to measure \nhlos as a function of time, thus deriving properties of the obscuring clouds through extensive, continuous monitoring campaigns \citep[e.g.][]{Markowitz2014}. This requires observing a full `eclipsing event'; that is, the ingress and egress of individual clouds into the l.o.s. However, very few X-ray instruments have the capability to provide this sort of continuous, daily monitoring over the necessary timescales (i.e. years to decades). In fact, the most extensive of such campaigns monitored 55 individual sources \citep{Markowitz2014}, spanning a total of 230 years of equivalent observing time with the Rossi X-ray Timing Explorer \citep[RXTE, ][]{Jahoda2006}. It resulted in the detection of only eight and four individual eclipsing events, in {Seyfert 1 and Seyfert 2} galaxies respectively. Despite the poor statistics, this study is the most complete to date, and has thus been used to calibrate one of the most recent X-ray emission models based on clumpy tori \citep{Buchner2019}.

Due to these observing difficulties, works focusing on comparing archival observations, or monitoring campaigns comprised of a few individual observations, are much more common. The $\Delta N_{\rm H,los}$ between two different observations separated by a given $\Delta t$ has been used to place upper limts on cloud distances to the SMBH in a number of works \citep[e.g.][]{Risaliti2002,Risaliti2005,Pizzetti2022,Marchesi2022}.

Only a handful of works exist that systematically analyze larger (i.e. $10-20$ sources) samples of AGN. Interestingly, works focusing on {Seyfert 2 galaxies} all reach a similar conclusion: less than half of nearby obscured AGN show \nhlos variability (7/20 sources in \citealt{Laha2020}, 11/25 in \citealt{HernandezGarcia2015}, 5/12 sources in \citealt{Torres-Alba2023}, 5/13 in \citealt{Pizzetti2024}). It is currently unclear if such a low fraction of \nhlos-variable sources is expected by clumpy torus models.

In order to further explore \nhlos variability in the local Universe, we have started an effort to analyse all archival data of AGN in the sample of \cite{Zhao2020} that have multiple soft X-ray observations. This sample was selected because it is comprised by obscured, yet Compton-thin AGN (i.e. \nhlos$=10^{23}-10^{24}$~cm$^{-2}$), making it possible to tightly constrain \nhlos, while also deriving torus geometrical properties via the usage of recent X-ray torus models \citep[see e.g. ][for details]{Zhao2020,Torres-Alba2023}. Furthermore, all sources contain at least one \nustar observation \citep[Nuclear Spectroscopic Telescope Array,][sensitive in the $3-78$~keV range]{Harrison2013}, which is necessary to derive the mentioned torus properties. Finally, all sources are detected by \textit{Swift}-BAT (Burst Alert Telescope, observing in the 15$-$150~keV range), implying their 15$-$150~keV flux is $>5\times 10^{-12}$~erg~s$^{-1}$~cm$^{-2}$, making it likely that archival soft X-ray observations will have high enough number of counts for modeling \nhlos variability\footnote{See \cite{Torres-Alba2023,Pizzetti2024} for further details on the sample selection}.

The first 25 sources within the sample, with the highest chance of being \nhlos-variable, were presented in \cite{Torres-Alba2023,Pizzetti2024}. It resulted in only 37\% being confidently classified as \nhlos-variable. As part of this effort, we have also developed a method to quickly flag potential \nhlos variability from hardness ratio comparisons of archival observations \citep{Cox2023}, and are currently working on applying it to the whole of the \chandra archive (Cox et al. in prep.). We have also presented the analysis of particularly interesting sources \citep[NGC 7974 and NGC 6300,][Sengupta et al. in prep., respectively]{Pizzetti2022}.

{Within this sample,} Mrk 477 was further selected for monitoring campaigns for two main reasons: it was already seen to be \nhlos-variable in \cite{Zhao2020}, as well as our preliminary analysis of the existent archival data; and it is the brightest source in the sample. In fact, Mrk 477 is the closest/brightest type II quasar, at a distance of 163 Mpc \citep[see e.g.][and references therein]{Heckman1997,Ramos-Almeida2023}, making it the perfect source for an inexpensive monitoring campaign. 

In this work, we analyze the existing archival data (five observations; two by \xmm, two by \nustar, and one by \chandra), as well as the data from two dedicated monitoring campaigns (five observations by \swift, and five more by \nustar). In Sect. \ref{data}, we describe the observations and the data reduction procedures. In Sect. \ref{analysis}, we describe the spectral analysis methodology and the X-ray torus reflection models used. In Sect. \ref{results}, we present spectra and best-fit parameters for the joint analysis of the 15 observations. In Sect. \ref{discussion} and \ref{conclusions}, we present our discussion and conclusions, respectively. 

\section{Observations and Data Reduction}\label{data}

Mrk 477 has been observed in the X-rays a total of 15 times, five of which are archival observations taken before any dedicated monitoring campaigns. Among those, there are two \xmm observations taken two days apart in 2010, followed by two \nustar observations taken 9 days apart in 2014. Finally, there is a \chandra observation taken in 2015.

Our first dedicated monitoring campaign took place between May and November 2021 (Proposal Number 1720100, Swift Cycle 17, PI: Torres-Alb\`a), and is comprised of 5 \swift observations of a total of $\sim$10~ks each, separated by consecutive time differences such that $\Delta t \simeq$ 2 weeks, 1 month, 2 months, 3 months (see Tables~\ref{tab:AllObservations} and~\ref{tab:SwiftCampaign} for exact details). The \nustar monitoring campaign took place between December 2022 and June 2023 (Cycle 8, PI: Torres-Alb\`a), with five $\sim20$~ks observations, following the same pattern of time differences between consecutive observations. A summary of observations can be found in Table~\ref{tab:AllObservations}.

\begin{deluxetable*}{ccccccc}
\tablecaption{Summary of all observations analyzed in this work.} 
\label{tab:AllObservations}
\tablehead{\colhead{Date} &  \colhead{Observatory} & \colhead{Obs ID} & \colhead{Exp. Time} & \colhead{Counts} &\colhead{$\Delta t$}  \\ \colhead{(1)} & \colhead{} & \colhead{(2)} & \colhead{(3)} & \colhead{(4)} & \colhead{(5)} }
\startdata
2010-07-21 & \xmm & 0651100301 & 7.1  & 2193 & $-$  \\
2010-07-23 & \xmm & 0651100401 & 6.5 & 2157 & 2  \\
2014-05-15  & \nustar  & 60061255002 & 18.1  & 4045 & 1392\\
2014-05-24 & \nustar & 60061255004 & 17.1  & 4604 & 9\\
2015-07-06  & \chandra & 17121 & 10.1  & 623 & 408\\
2021-05-01 & \swift & 00095968001 & 7.9 & 98 & 2126\\
2021-05-15 & \swift & Merged 1 & 10.6  & 157 & 14\\
2021-06-15 & \swift & Merged  2& 8.2 & 100 & 31\\
2021-08-15  & \swift & 00095968008 & 8.3  & 110 & 61\\
2021-11-15 & \swift & 00095968009 & 8.8 & 89 & 92\\
2022-12-13 & \nustar & 60802020002 & 22.1 & 4291 & 393\\
2022-12-31 & \nustar & 60802020004 & 22.7 & 3982 & 18\\
2023-02-01 & \nustar & 60802020006 & 21.6 & 3357 & 32\\
2023-03-26 & \nustar & 60802020008 & 22.1 & 4015 & 53\\
2023-06-28 & \nustar & 60802020010& 22.0 & 2791 & 94\\
\enddata
\tablenotetext{}{\textbf{Notes:} (1): Observation date. (2): Observation ID. Two of the \swift observations have been split into multiple observations, taken in concsecutive days. More specific details on these (including Obs. ID) can be found in Table \ref{tab:SwiftCampaign}. (3): Net exposure time in units of ks. For \xmm, we list the pn camera exposure. (4): Total counts in the 0.5$-$7 keV band (for \swift,\xmm and \chandra) or in the 3$-$50 keV band for \nustar. Counts from different cameras/modules are summed up. (5): Time difference between this observation and the previous observation of the source, in days.}
\end{deluxetable*}

All spectra were binned with at least 25 counts per bin in order to use $\chi^2$ statistics for the fitting, with the exception of the \swift data, which did not have a large enough number of counts. The \swift data was thus binned with 5 counts per bin, and fit using W-stat instead. We fit the spectra using XSPEC v.12.11.1 \citep[][in HEASOFT version 6.28]{Arnaud1996}. We took into account Galactic absorption in the line of sight, according to the values measured by \cite{Kalberla2005}, as well as the \cite{Verner1996} photoelectric cross-section. We fixed metal abundances to the solar value, using the \cite{Anders1989} cosmic abundance measurements. 
All luminosity distances are computed assuming a cosmology with H$_0$=70~km s$^{-1}$ Mpc$^{-1}$, and $\Omega_\Lambda$=0.73.

\subsection{XMM-Newton Data Analisys}

We used the \xmm Science Analysis Software (XMM SAS) v20.0.0 to reduce the \xmm data from both MOS1 and 2 as well as the PN camera \citep{Jansen2001}, adopting the standard procedure, cleaning for flaring periods. Once the event file was cleaned, we extracted the source spectra from a 30'' circular region centered on the source. The background spectra were obtained using a circle of radius 45'', located near the source (avoiding nearby objects as well as CCD lines). 

\subsection{NuSTAR Data Analisys}

The \nustar data was retrieved from both NuSTAR Focal Plane Modules \citep[FPMA/B;][]{Harrison2013} and processed using the NuSTAR Data Analysis Software (NUSTARDAS)
v2.1.2. The Calibration Database (CALDB) v.20211020 was run to calibrate the event data files through the \texttt{nupipeline} task. We extracted the source and background spectra using the \texttt{nuproducts} script. For both FPMA and B we used a circular region with a 50'' radius, centered around the source, to extract its spectrum. For the background, we adopted an annulus (inner radius 100'', outer radius 160'') surrounding the source.

\subsection{Chandra Data Analisys}

The \chandra data was reduced with the Chandra Interactive Analysis of Observations software (CIAO) v4.14 \citep{Fruscione2006}. To extract the source spectra, a circular region of radius 5'' was used, centered arund the source. For the background, we used an annulus (inner radius 6'', outer radius 15'') surrounding the source, making sure any additional resolved sources are excluded. 

\subsection{Swift-XRT Data Analisys}

\begin{figure}
    \centering
    \hspace{-0.65cm}
    \includegraphics[scale=0.5]{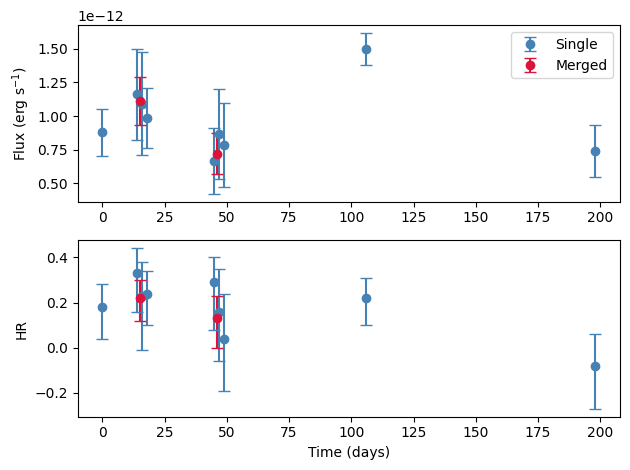}
    \caption{Flux (top) and Hardness Ratio (bottom) as a function of time for the \swift campaign, showing all observations taken, as described in Table \ref{tab:SwiftCampaign}. Data groups 2 and 3 are shown as three single observations each (in blue), and as the merged combination of all three (in crimson). Data groups 1, 4 and 5 were originally taken as one single observation, and thus are shown in blue.}
    \label{fig:flux_var}
\end{figure}

The \swift data \citep{Burrows2005} were reduced using the XRT Data Analysis Software (XRTDAS) v.3.6.1. The Calibration Database (CALDB) v.20211020 was run to calibrate the event data files through the \texttt{xrtpipeline} task. We extracted the source spectra using a circular region with a 20'' radius, while the background was extracted using an annulus (inner radius 30'' and outer radius 60''), placed around the source, making sure to exclude any additional resolved sources. Two of the 5 observations belonging to the \swift monitoring campaign were split into three shorter observations, taking place within consecutive days, which are detailed in Table \ref{tab:SwiftCampaign}. The shorter observations do not have a sufficiently large number of counts to consider them individually for spectral fitting, and thus we use the `merged' observations for that purpose. To ascertain it is safe to merge the observations (i.e. that no obvious large variability has taken place between them) we measured the flux and hardness ratio (HR$=$(H$-$S)/(H+S); where H and S are the number of counts in the 2$-$7~keV band and the 0.5$-$2~keV band, respectively). These results are shown in Table \ref{tab:SwiftCampaign} and Fig. \ref{fig:flux_var}. As can be appreciated, no significant changes are taking place within the single observations, and thus we merge them. 

\begin{deluxetable*}{ccccccc}
\tablecaption{\swift-XRT campaign observation details} 
\label{tab:SwiftCampaign}
\tablehead{\colhead{Set} &  \colhead{Date} & \colhead{Obs ID} & \colhead{Exp. Time} & \colhead{Counts} &\colhead{Flux} & \colhead{HR} \\ \colhead{} & \colhead{} & \colhead{} & \colhead{[ks]} & \colhead{} & \colhead{[$10^{-13}$ erg $s^{-1}$]} & \colhead{} \\ \colhead{(1)} & \colhead{(2)} & \colhead{(3)} & \colhead{(4)} & \colhead{(5)} & \colhead{(6)} & \colhead{(7)} }
\startdata
1 & May 1 2021 & 00095968001 & 7.9  & 98 & 8.8$\pm$1.7 & 0.18$^{+0.10}_{-0.14}$ \\
2 & May 15 2021 & 00095968002 & 3.9& 48 & 11.6$\pm$3.4 & 0.33$^{+0.11}_{-0.17}$ \\
 & May 17 2021 & 00095968003 & 1.9  & 34 & 10.9$\pm$3.8 & 0.23$^{+0.15}_{-0.24}$\\
 & May 19 2021 & 00095968004 & 4.8  & 73 & 9.9$\pm$2.2 & 0.24$^{+0.10}_{-0.14}$\\
  & Merged 1& $-$ & 10.6  & 155 & 11.1$\pm$1.8 & 0.22$^{+0.08}_{-0.10}$\\
3 & June 15 2021 & 00095968005 & 3.3 & 33 & 6.7$\pm$2.5 & 0.39$^{+0.11}_{-0.21}$\\
 & June 17 2021 & 00095968006 & 1.4  & 24 &8.9$\pm$4.9 & 0.16$^{+0.19}_{-0.35}$\\
 & June 19 2021 & 00095968007 & 3.5 & 39 & 9.7$\pm$3.1 & -0.02$^{+0.20}_{-0.31}$\\
  & Merged 2& $-$ & 8.2  & 96 & 7.2$\pm$1.5 & 0.13$^{+0.10}_{-0.13}$\\
4 & August 15 2021 & 00095968008 & 8.3 & 110 & 14.9$\pm$1.2 & 0.22$^{+0.09}_{-0.12}$\\
5 & November 15 2021 & 00095968009 & 8.8 & 89 & 7.4$\pm$1.9 & -0.08$^{+0.14}_{-0.19}$\\
\enddata
\tablenotetext{}{\textbf{Notes:} (1): Observation group. (2): Individual observation date. (3): Observation ID. (4): Exposure time. (5): Total counts in the 0.5$-$7 keV band. (6): Observed flux in the 0.5$-$7 keV band. (7): Hardness ratio as (H-S)/(H+S), with S being the number counts in the 0.5$-$2 keV band, and H being the numbers counts in the 2$-$7 keV band.}
\end{deluxetable*}

\section{X-ray Spectral Analysis}\label{analysis}

We fit the source using two recently-developed, self-consistent torus models: the broadly-used, homogeneous torus model \bor \citep{Balokovic2018}; and the clumpy torus model \uxc \citep{Buchner2019}. This choice is consistent with the approach used in previous works by our group \citep{Pizzetti2022,Torres-Alba2023,Pizzetti2024}, which used the same models to study N$_{H,los}$ variability in a sample totaling 27 sources so far. Mrk 477 is drawn from the same parent sample, as detailed in Sect. \ref{intro}. Using the same models allows us to derive geometric parameters for Mrk 477 that we can compare to the rest of the sample, which are model specific. More specifically, with \bor we measure: torus covering factor (\cf), and average torus column density (\nhav).  With \uxc, we instead derive the width of the Gaussian cloud distribution (\sigtor), and the covering factor of a (possible) additional Compton-thick reflector (\ctk). Both models, of course, allow us to obtain values of the line-of-sight column density (\nhlos), which is the objective of this monitoring campaign; as well as values for the inclination angle of the source (\iang). 


The mentioned geometric parameters are derived from the fitting of the reflection component, which is typically assumed to originate in the torus. Because of this, and as already done for the rest of the sample \citep{Pizzetti2022,Torres-Alba2023,Pizzetti2024}, we consider that the parameters defining this reflected emission do not change across time. That is, the torus as a whole does not vary its properties significantly on timescales of a few decades. Therefore, we fit all observations together, imposing that all reflection parameters remain tied across observations, but allowing intrinsic luminosity variability as well as \nhlos variability across different observations. As done in the mentioned works, we also assume that the photon index, $\Gamma$, of the main powerlaw emission does not vary. We further explore the reliability of these assumptions in Sect. \ref{results}.

The source is fit with the following setup for both torus models:
\begin{equation}
    Model = phabs * ( apec + zgauss + C_{\rm norm}*AGNModel),
\end{equation}
where \texttt{phabs} accounts for Galactic absorption in the line of sight, \texttt{apec} accounts for thermal Bremmstrahlung emission originated in hot gas within the host galaxy, C$_{\rm norm}$ is a constant that accounts for intrinsic luminosity variability between different observations, and the AGN model setup is defined in Sects. \ref{borus} and \ref{uxclumpy} for \bor and \uxc, respectively. An additional emission line component ($kT\sim0.3$~keV), modeled with \texttt{zgauss}, was added to adequately reproduce the soft band ($<$2 keV) emission.

\begin{figure*}[t!]
    \centering
    \hspace{-0.65cm}
    \includegraphics[scale=0.54]{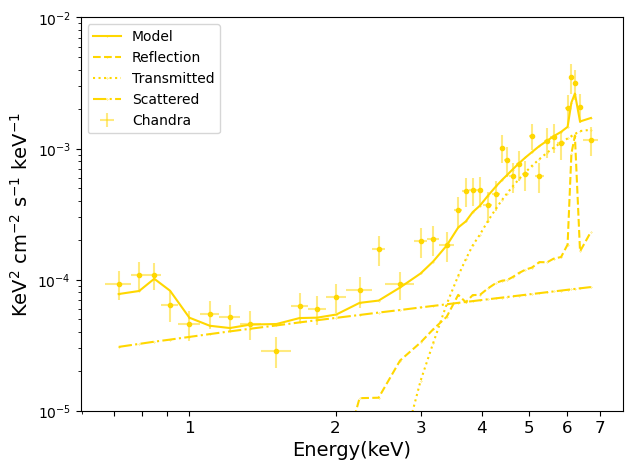}
    \includegraphics[scale=0.54]{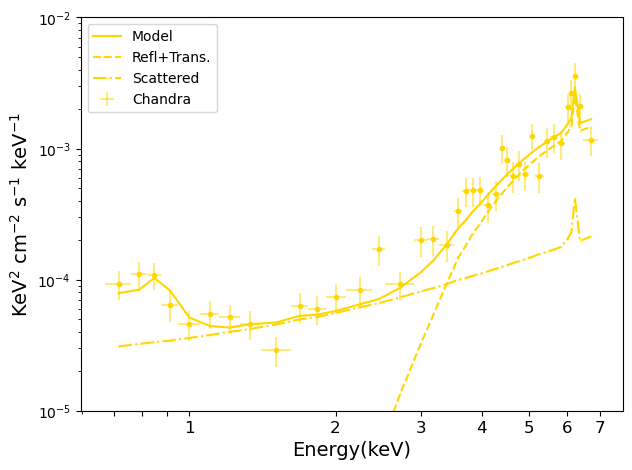}
    \includegraphics[scale=0.54]{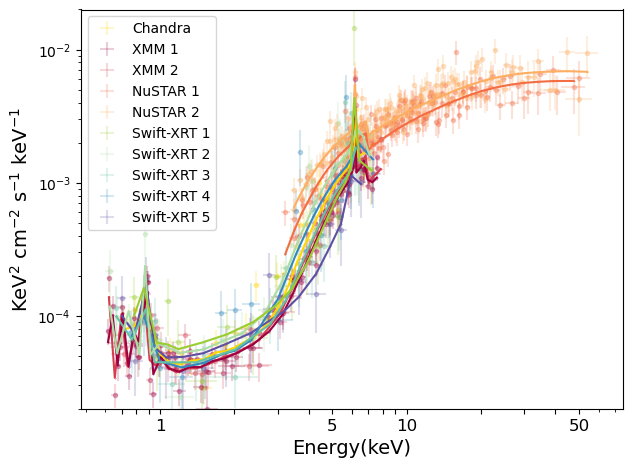}
    \includegraphics[scale=0.54]{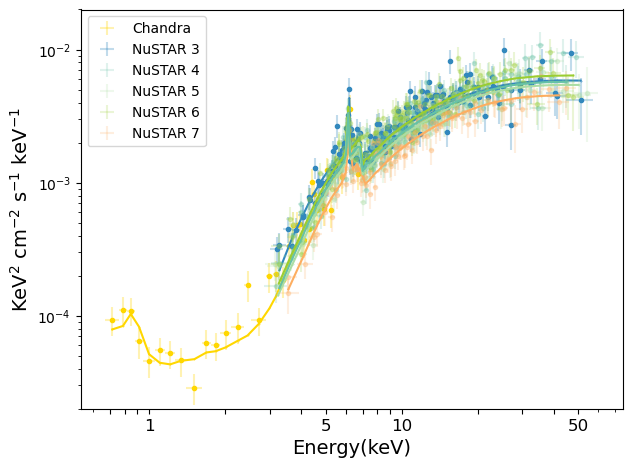} 
    \caption{\textbf{Top:} Chandra spectrum of Mrk 477, one of the 15 observations fit simultaneously in this work. \textit{Top Left:} \bor fit to the spectrum, showing the subcomponents listed in Eq. 2 (reflected, transmitted or line-of-sight, and scattered emission). \textit{Top Right:} \uxc fit to the spectrum, showing the subcomponents listed in Eq. 3 (where the reflected and transmitted components are one single table, with the scattering component added apart). {Bottom: All spectra of Mrk 477 used in this work, separated into two panels for display purposes. The \textit{Chandra} spectrum is shown twice, to serve as a comparison point against the \textit{NuSTAR} campaign shown on the right plot. The best-fit model using \uxc is shown overlaid on the data with solid lines, to better highlight the spectral variability. All individual spectra and their fit with both \bor and \uxc can be found in Appendix \ref{App:spectra}.}}
    \label{fig:chandra_spectra}
\end{figure*}

\subsection{\bor}\label{borus}

\bor \citep{Balokovic2018} is a uniform torus model, which consists of a cold material reflection component (i.e. reflection from the torus), accounting for both continuum and lines. The geometry of this reflecting material can be changed via a covering factor, \cf (\cf $\in$[0.1,1.0]), and its average column density can be changed via \nhav. In ``decoupled'' configuration, one can disentangle the line-of-sight column density (or obscuration, \nhlos) from \nhav, thus obtaining a first approximation to a clumpy medium. That is, the density through which we look at the source is different from the average density of the material, even if the emission reflected from the torus is modeled with a homogeneous medium. The model also includes an inclination angle of the torus, \iang (\iang $\in [18^\circ-87^\circ]$). As explained above, we impose that \cf, \nhav, and \iang do not vary across different observations. 

Finally, the model also includes a high-energy cutoff, $E_{\rm cut}$ \citep[which we froze at 300 keV, as per the results of studies on the local obscured AGN population; ][]{Balokovic2020}, and an iron abundance, which we froze at Solar abundance due to our inability to constrain it given the data. 

The AGN model setup for \bor is as follows:
\begin{equation}
\begin{split}
    AGN Model = borus02\_v170323a.fits \  + \\
    zphabs*cabs*cutoffpl \ + \  F_s * cutoffpl
\end{split}
\end{equation}
where the table accounts for the reflected emission as described above, and \texttt{zphabs} and \texttt{cabs} are photoelectric absorption and Compton scattering away from the line of sight due to obscuring material (affecting the main powerlaw, \texttt{cutoffpl}, originating from the corona). \fscat is the scattering fraction, usually on the order of a few percent, which accounts for the intrinsic powerlaw emission of the AGN that either interacts elastically with the torus. Fig. \ref{fig:chandra_spectra} shows this decomposition applied to one of the spectra fit in this work.

{We note that \bor has been shown to have a problem in the convolution of the Green functions, which can result in an erroneously generated spectrum at both low and high energies \citep{vanderMeulen2023}. At high energies, discrepancies start to be apparent at $E>40$~keV, and become more significant as energy increases (see Fig. 16 in the aforementioned work). We note that the effect of this problem in our work is minimal, given how we only fit the spectra up to $\sim 50$~keV, and the errors in the data in the $40-50$~keV range are large enough to account for the discrepancy. At low energies, the problem only occurs when not observing through the torus (i.e. Cos(\iang)$\gg 0$), which does not affect this work since the \bor tables used are those for Cos(\iang)$=0$. We also note that the determination of \nhlos by \bor has always been in good agreement with both the \uxc and \myt values in our previous work \citep[][]{Marchesi2019b,Pizzetti2022,Pizzetti2024,Torres-Alba2023,Sengupta2024}, which now encompasses more than 150 individual observations. As such, our \nhlos values are unlikely to be affected by this issue.}

However, as they become available, we recommend the usage of alternative torus models {\citep[e.g. XCLUMPY, SKIRT or RXTorusD,][respectively]{tanimoto2019,vanderMeulen2023,Ricci2023}, particularly when fitting data up to higher energy ranges.}

\subsection{\uxc}\label{uxclumpy}

\uxc is a clumpy torus model \citep{Buchner2019}, based on the work of \cite{Nenkova2008}, which includes a cloud distribution calibrated using known AGN column density distributions \citep{Aird2015,Buchner2015,Ricci2015} and frequencies of observed eclipsing events \citep{Markowitz2014}. The clouds are set in a Gaussian distribution, above and below the equatorial plane, with a characteristic width \sigtor (\sigtor $\in [6^\circ-90^\circ]$). This distribution is viewed from a certain inclination angle, \iang (\iang $\in [0\degree-90\degree]$).

Unlike \bor, the model can be setup using a single component, which accounts for the reflected continuum and lines, as well as the absorbed main powerlaw emission. Like \bor, the model includes a high-energy cutoff, which we again froze to $E_{\rm cut}=300$~keV. 

\urldef{\footurl}\url{https://github.com/JohannesBuchner/xars/blob/master/doc/uxclumpy.rst}

In addition to the cloud distribution, the model includes a ring of Compton-thick material, which is a necessary addition when fitting sources with particularly strong reflection \cite{Buchner2019}\footnote{See \footurl for a representation of different possible geometries}. This material is characterized by a covering factor, \ctk (\ctk $\in [0-0.6]$), which takes the value of \ctk=0 when the source does not require a strong reflection component to explain its spectrum. As explained above, we impose that \ctk, \sigtor, and \iang do not vary across different observations. 

The AGN model setup for \uxc is as follows:
\begin{equation}
\begin{split}
    AGN Model = uxclumpy. fits + \\
 F_s * uxclumpy\_scattered.fits,
\end{split}
\end{equation}
where the first table includes both line-of-sight and reflected emission, and \fscat is the scattering fraction. \uxc provides a scattered component with a correction for emission that may leak from the torus after being reflected by the torus at least once.  Fig. \ref{fig:chandra_spectra} shows this decomposition applied to one of the spectra fit in this work.

\section{Results}\label{results}

In this section we present the results of the fitting process described in Sect. \ref{analysis}. Table \ref{tab:reflection} lists the common parameters for all observations (i.e. soft emission parameters, torus parameters, photon index, and scattering fraction); while Table \ref{tab:nhvar} lists the values that are set free to vary for each observation (i.e. \nhlos, \cnorm), {described in more detail} in Sect. \ref{obscuration_variability}. 

Table \ref{tab:reflection} shows the results for the best-fit values of the reflection parameters for three different datasets: the archival data plus the \swift monitoring campaign, the \nustar monitoring campaign, and the full dataset. This is done for both models, \bor and \uxc, with the objective of testing the methodology described in Sect. \ref{analysis}. That is, the hypothesis that we can assume the reflection parameters do not vary on the timescales considered in this work ($\sim~10$~yr). As can be seen in Table \ref{tab:reflection}, there is good agreement between all three datasets {(see Appendix \ref{App:reflection} for a more thorough discussion on the viability of this hypothesis)}.

Despite the strong agreement between the different datasets, the \bor and \uxc results present significant difference when it comes to the determination of photon index, $\Gamma$. This is a common occurrence when using different torus models \citep[see e.g.][]{Torres-Alba2023,Pizzetti2024}, likely due to the degeneracy between the shape of the reflector (different for each model) and the powerlaw slope of the incident emission. 

As for the rest of the parameters, \bor favors a high-covering factor reflector, but with low obscuring column density (\nhav$\sim 8 \times 10^{22}$ cm~$^{-2}$). This is in agreement with the \uxc results, for which the covering factor of the thick reflector (\ctk) was stuck at the lowest bound, and thus we froze it to zero\footnote{As suggested per the model instructions.}. The cloud coverage around the torus, \sigtor, is at its highest limit, in accordance with the high-covering factor found by \bor. Thus, the models qualitatively agree that the AGN in Mrk 477 is surrounded by a reflector with broad coverage, but low density. 

{\uxc does not provide a direct estimate of the average torus column desnsity, but it is possible to estimate the equatorial column density (N$_{\rm H,eq}$), as detailed in \cite{Pizzetti2024}. The computation of N$_{\rm H,eq}$ depends solely on the values of C$_{\rm TK}$ and \sigtor. Using the parameters listed in Table \ref{tab:reflection}, we obtain N$_{\rm H,eq}\sim 6 \times 10^{23}$~cm$^{-2}$.}

The inclination angle is fully unconstrained by both models, and the discrepancy between \fscat values is due to the different parameterization of the models, and is a common occurrence for the rest of the parent sample as well \citep[][]{Torres-Alba2023}.

\begin{deluxetable*}{c|cc|cc|cc}\label{tab:reflection}
\tablecaption{Comparison of common parameters for the three fitting datasets: archival + \swift, \nustar-only, all.}
\tablehead{\colhead{} & \colhead{borus} & \colhead{borus Cfree} & \colhead{UXCLUMPY} & \colhead{borus} & \multicolumn{2}{c}{NuSTAR only}\\ 
}
\tablehead{\colhead{} & \multicolumn{2}{c}{Archival+XRT} & \multicolumn{2}{c}{NuSTAR only} &  \multicolumn{2}{c}{Full dataset}\\
&\bor & \uxc & \bor & \uxc & \bor & \uxc }
\startdata
kT  & 0.29$^{+0.06}_{-0.05}$ & 0.30$^{+0.06}_{-0.05}$ & $-$ & $-$ & 0.29$^{+0.07}_{-0.06}$ &  0.29$^{+0.06}_{-0.05}$ \\
E$_{\rm line}$ & 0.90$^{+0.01}_{-0.01}$ & 0.91$^{+0.01}_{-0.01}$ & $-$ & $-$ & 0.91$^{+0.01}_{-0.01}$ & 0.91$^{+0.01}_{-0.01}$  \\
$\Gamma$ & 1.54$^{+0.05}_{-0.05}$  & 1.77$^{+0.06}_{-0.05}$ & 1.57$^{+0.04}_{-0.05}$ & 1.80$^{+0.03}_{-0.05}$  & 1.54$^{+0.05}_{-0.06}$ & 1.78$^{+0.03}_{-0.03}$ \\
$N_{H,av}$ & 0.08$^{+0.02}_{-0.02}$ & $-$ & 0.08$^{+0.02}_{-0.02}$ & $-$ & 0.08$^{+0.02}_{-0.02}$ & $-$\\
\cf  & 1.00$^{+u}_{-0.12}$ & $-$ & 1.00$^{+u}_{-u}$ & $-$ & 1.00$^{+u}_{-0.08}$ & $-$\\
\ctk  & $-$ & 0*  & $-$ & 0* & $-$& 0*\\
$\sigma_{\rm tor}$ & $-$ & 84.0$^{+u}_{-15.6}$ & $-$ & 27.8$^{+56.2}_{-1.8}$ & $-$ & 84.0$^{+u}_{-7.3}$\\
Cos ($\theta_{Obs}$) & 0.05$^{+u}_{-u}$ & 0.00$^{+u}_{-u}$ & 0.05$^{+u}_{-u}$ & 0.00$^{+u}_{-u}$ & 0.05$^{+u}_{-u}$ & 0.49$^{+u}_{-u}$ \\ 
F$_s$ (10$^{-2}$) & 2.46$^{+1.23}_{-0.46}$  & 18.9$^{+2.9}_{-3.2}$ & 6.53$^{+0.}_{-0.}$ & 16.1$^{+9.2}_{-1.8}$ & 2.80$^{+0.83}_{-0.51}$ & 20.6$^{+1.9}_{-3.3}$ \\
norm (10$^{-3}$) & 1.51$^{+0.37}_{-0.54}$  & 2.38$^{+0.50}_{-0.51}$ & 1.62$^{+0.19}_{-0.31}$& 3.26$^{+0.11}_{-0.25}$ &1.34$^{+0.32}_{-0.35}$ & 2.14$^{+0.35}_{-0.26}$\\
\enddata
\begin{tablenotes}{\footnotesize \textbf{Notes:} $kT:$ \texttt{apec} model temperature, in units of keV. $E_{\rm line}$: Central energy of the added Gaussian line, in units of keV. $\Gamma$: Powerlaw photon index. $N_{H,av}$: Average torus column density, in units of $10^{24}$ cm$^{-2}$. $C_{\rm f}$: Covering factor of the torus. \ctk: Covering factor of the additional Compton-thick ring reflector. \sigtor: width of the Gaussian cloud distribution. cos ($\theta_{i}$): cosine of the inclination angle. cos ($\theta_{i}$)=1 represents a face-on scenario. F$_s$: Fraction of scattered continuum. Norm: Normalization of the AGN emission. (*) Denotes a parameter frozen to the shown value. ($-$) Denotes a parameter that is not included in a given model and/or dataset. We note that the \nustar-only dataset does not include a soft X-ray model, given how the data starts at 3 keV. ($-u$) refers to a parameter being compatible with the hard limit of the available range.} 
\end{tablenotes}
\end{deluxetable*}

Table \ref{tab:nhvar} shows the obscuration for each observation, for the full dataset, in two different scenarios: with and without allowing intrinsic flux variability. There are no significant differences, either between these two scenarios, or between the two used models; with all agreeing the source varies in \nhlos between a few epochs. A more thorough discussion can be found in Sect. \ref{obscuration_variability}.

The spectra, while fit simultaneously, are shown {in two separate plots}, to avoid overcrowding. Fig. \ref{fig:chandra_spectra} (top) shows the \bor and \uxc fits, divided into its multiple subcomponents, applied to the \chandra observation alone. {The bottom panel shows all spectra used in this work, and their best-fit with the \uxc model, to highlight the spectral shape diversity. All individual observations, along with their best-fit models with \bor and \uxc,} can be found in Appendix \ref{App:spectra} (Fig. \ref{fig:spectra_borus} and \ref{fig:spectra_uxc}). Given how each spectrum is shown separately, each panel also contains the fit to the \chandra spectrum (as shown in Fig. \ref{fig:chandra_spectra}), to make it easier to visualize the variability.

\section{Discussion}\label{discussion}

In this section, we discuss the main results of the analysis, regarding the variability of \nhlos and its characteristics. {Section \ref{obscuration_variability} discusses obscuration variability for Mrk~477, while Sect. \ref{nhlosvar_fullsample} examines the probability of variability for the whole sample of sources analyzed by our group so far. Additionally, Sect. \ref{blr_var} discusses the viability of obscuration variability originating in the BLR. Section \ref{other_works} discusses our results in the context of previous works.

\subsection{Obscuration variability}\label{obscuration_variability}

There are 15 individual observations available for Mrk 477 (see Table \ref{tab:AllObservations}), which were all used in the `full dataset' fit shown in this section. The timespan between consecutive observations is as small as 2 days and as large as $\sim6$~yr. This results in a large variety of timescales ($\Delta t$) for which to explore \nhlos variability. 

Table \ref{tab:nhvar} shows the result of the \nhlos variability analysis for the full dataset. {We consider intrinsic flux variability, which is accounted for through a cross-normalization constant, labeled $C_{\rm telescope,number}$, according to the number observation taken with each telescope, in chronological order. All fluxes are normalized to the \textit{Chandra} observation.} The same system is used for the \nhlos values listed in both scenarios.

\begin{deluxetable}{c|c|c}
\label{tab:nhvar}
\tablecaption{Mrk 477, variability study for the full campaign.}
\tablehead{\colhead{Model} & \colhead{\bor} & \colhead{\uxc} \\  }
\startdata
red $\chi^2$  & 1.00  & 1.01 \\
$\chi^2$/d.o.f.  & 1277.7/1272  &  1286.2/1270\\
\hline 
$N_{H,xmm1}$ & 0.40$^{+0.04}_{-0.04}$  & 0.35$^{+0.03}_{-0.03}$  \\
$N_{H,xmm2}$  & 0.36$^{+0.04}_{-0.03}$ & 0.32$^{+0.03}_{-0.04}$  \\ 
$N_{H,nus1}$  & 0.21$^{+0.03}_{-0.03}$  & 0.21$^{+0.03}_{-0.03}$ \\
$N_{H,nus2}$  & 0.17$^{+0.02}_{-0.02}$  & 0.17$^{+0.02}_{-0.03}$ \\
$N_{H,Ch}$ & 0.34$^{+0.06}_{-0.05}$  & 0.28$^{+0.04}_{-0.05}$  \\
$N_{H,swift1}$ & 0.54$^{+0.23}_{-0.15}$  & 0.49$^{+0.24}_{-0.16}$\\
$N_{H,swift2}$ & 0.30$^{+0.09}_{-0.07}$  & 0.26$^{+0.09}_{-0.06}$\\
$N_{H,swift3}$ & 0.37$^{+0.12}_{-0.10}$  & 0.33$^{+0.15}_{-0.11}$\\
$N_{H,swift4}$ & 0.30$^{+0.12}_{-0.08}$  & 0.25$^{+0.12}_{-0.08}$\\
$N_{H,swift5}$ & 0.75$^{+0.55}_{-0.20}$  & 0.71$^{+0.71}_{-0.25}$ \\
$N_{H,nus3}$ & 0.29$^{+0.03}_{-0.04}$ & 0.27$^{+0.03}_{-0.03}$ \\
$N_{H,nus4}$ & 0.38$^{+0.05}_{-0.04}$ & 0.35$^{+0.03}_{-0.03}$ \\
$N_{H,nus5}$ & 0.34$^{+0.06}_{-0.05}$ & 0.37$^{+0.05}_{-0.04}$ \\
$N_{H,nus6}$ & 0.37$^{+0.05}_{-0.05}$ & 0.35$^{+0.04}_{-0.04}$ \\
$N_{H,nus7}$ & 0.43$^{+0.07}_{-0.06}$ & 0.39$^{+0.06}_{-0.05}$ \\\hline
$C_{xmm1}$ & 0.90$^{+0.13}_{-0.11}$ & 0.94$^{+0.13}_{-0.14}$  \\
$C_{xmm2}$ & 0.90$^{+0.13}_{-0.11}$ &0.94$^{+0.13}_{-0.14}$  \\ 
$C_{nus1}$  & 1.02$^{+0.20}_{-0.17}$ & 1.13$^{+0.30}_{-0.11}$  \\
$C_{nus2}$ & 1.15$^{+0.22}_{-0.18}$ & 1.42$^{+0.21}_{-0.24}$ \\
$C_{Ch}$  & 1* & 1*  \\
$C_{swift1}$  & 1.32$^{+0.40}_{-0.32}$ & 1.39$^{+0.40}_{-0.35}$\\
$C_{swift2}$ & 1.20$^{+0.33}_{-0.25}$ & 1.26$^{+0.33}_{-0.26}$\\
$C_{swift3}$  & 0.97$^{+0.31}_{-0.25}$ & 1.02$^{+0.32}_{-0.27}$ \\
$C_{swift4}$  & 0.97$^{+0.35}_{-0.27}$ & 1.00$^{+0.34}_{-0.26}$ \\
$C_{swift5}$  & 1.11$^{+0.31}_{-0.24}$ & 1.18$^{+0.32}_{-0.28}$ \\
$C_{nus3}$  & 1.08$^{+0.22}_{-0.18}$ & 1.27$^{+0.19}_{-0.22}$ \\
$C_{nus4}$ & 1.13$^{+0.26}_{-0.21}$ & 1.28$^{+0.19}_{-0.22}$  \\
$C_{nus5}$  & 1.09$^{+0.25}_{-0.20}$ & 1.22$^{+0.19}_{-0.22}$ \\
$C_{nus6}$ & 1.25$^{+0.27}_{-0.21}$ & 1.42$^{+0.21}_{-0.26}$ \\
$N_{H,nus7}$ & 0.92$^{+0.21}_{-0.17}$ & 1.03$^{+0.18}_{-0.19}$ 
\enddata
\begin{tablenotes}{\footnotesize \textbf{Notes:} red $\chi^2$ (or Stat): reduced $\chi^2$ or total Statistic. $\chi^2$(or Stat)/d.o.f.: $\chi^2$ (or total Statistic) over degrees of freedom. $N_{\rm H, inst., num.}$: Line-of-sight hydrogen column density for a given observation, in units of $10^{24}$ cm$^{-2}$. C$_{\rm inst., num.}$: Cross-normalization constant for a given observation, with respect to the intrinsic flux of the Chandra observation. (*) Denotes a parameter frozen to the shown value.}
\end{tablenotes}
\end{deluxetable}

{We note that, although the vast majority of cross-normalization constants are compatible with 1 (i.e. no intrinsic flux variability) at a 90\% confidence level (see Table \ref{tab:nhvar}), allowing flux variability still resulted in a better fit. The addition of intrinsic flux variability did not result in large changes in the accompanying \nhlos values for each observation, but did imply a slight increase in \nhlos errors. }

\begin{figure*}
    \centering
    \includegraphics[scale=0.5]{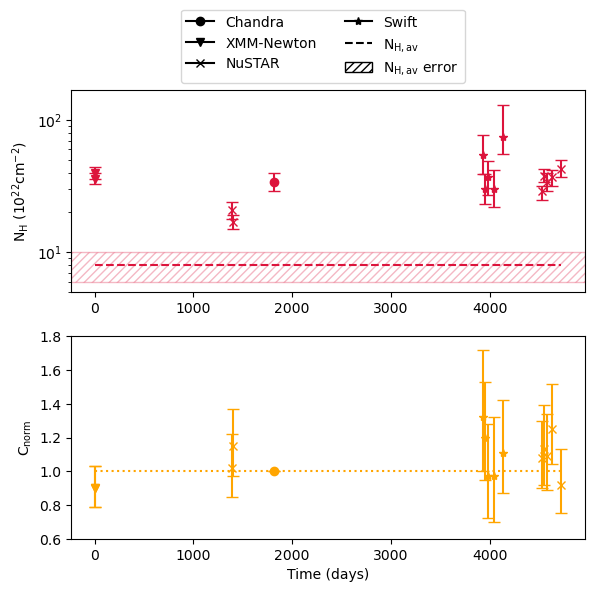}
    \includegraphics[scale=0.5]{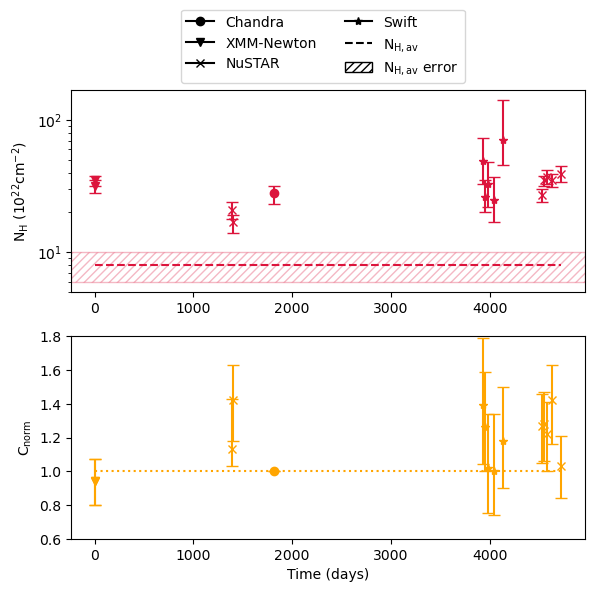} 
    \caption{\nhlos and $C_{\rm norm}$ variability as a function of time, for \bor (left) and \uxc (right). The \nhlos is compared to \nhav (shaded area) as determined by \bor in all panels.}
    \label{fig:nh_var}
\end{figure*}

The values of \nhlos versus time can be seen in Fig. \ref{fig:nh_var} for both models. The figure also includes the cross normalization variability.

{The archival datasets (e.g. the \xmm observations against the second \nustar observation) and the monitoring campaigns (e.g. second and last \swift observations, first and last \nustar observations) show \nhlos variability.} The agreement between \bor and \uxc is remarkable, with some minor differences. A few of the \nhlos values for \uxc are slightly lower, while $C_{\rm norm}$ values are higher (and incompatible with 1 at the 90\% confidence level). Such differences between combinations of \nhlos and $C_{\rm norm}$ between \bor and \uxc have been observed before \citep[e.g.][]{Torres-Alba2023,Pizzetti2024}, and tend to be more frequent when dealing with \nustar data. Looking at the $C_{\rm norm}$ values in Table \ref{tab:nhvar}, this is the case in this work as well.

Given the difficulty in visualizing the \nhlos variability at all possible timescales, we provide an additional representation. Fig. \ref{fig:nh_contours_borus} and \ref{fig:nh_contours_uxc} show 90\% confidence contours of \nhlos versus \cnorm for the full dataset fit, for \bor and \uxc respectively. 

\begin{figure*}
    \centering
    \hspace{-0.65cm}
    \includegraphics[scale=0.55]{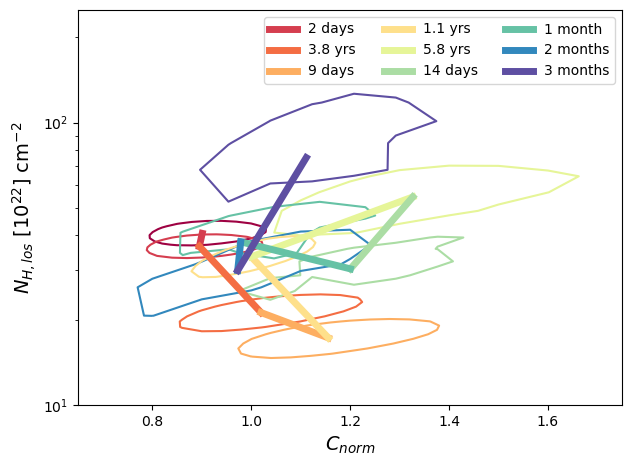}
    \includegraphics[scale=0.55]{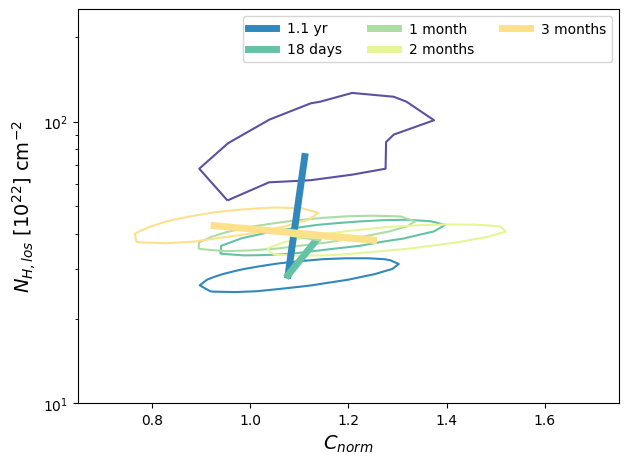}
    \caption{90\% confidence contours of \nhlos versus $C_{\rm norm}$ for the whole dataset, as fit by \bor. The thick, colored lines go from the best-fit value of one observation into the the best-fit value of the following observation, in chronological order. The legend shows the $\Delta t$ between the corresponding consecutive observations. These contours correspond the the full dataset shown in Table \ref{tab:nhvar}, and has simply been split into two panels to avoid overcrowding the plot. The left panel shows the archival data and \swift campaign, with observations going from red to dark blue in chronological order. The right panel starts with the last \swift observation, in dark blue, and goes from dark blue to yellow in chronological order.}
    \label{fig:nh_contours_borus}
\end{figure*}

\begin{figure*}
    \centering
    \hspace{-0.65cm}
    \includegraphics[scale=0.55]{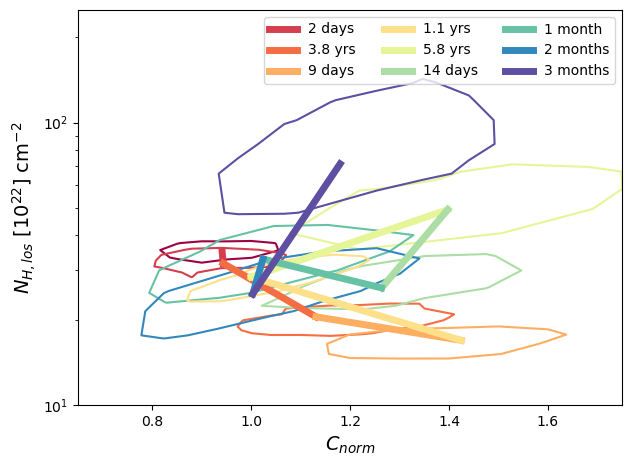}
    \includegraphics[scale=0.55]{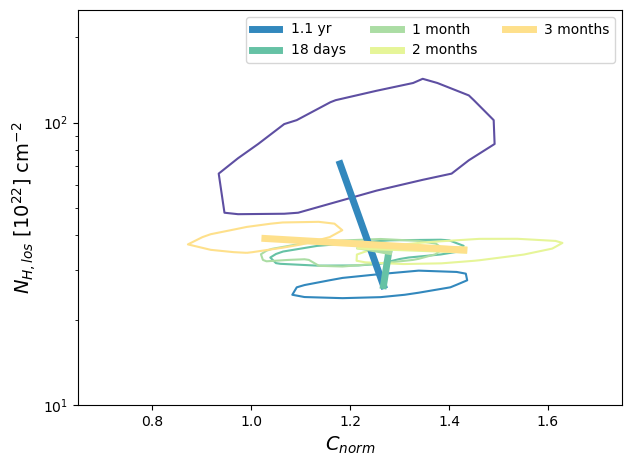}
    \caption{Same as Fig. \ref{fig:nh_contours_borus}, for \uxc.}
    \label{fig:nh_contours_uxc}
\end{figure*}

For each model, the data is split into two panels, which does not indicate that it corresponds to different datasets being fit separately. Rather, the split is needed in order to avoid overcrowding the plot. To make comparison easier, the scale is kept exactly the same in all four panels. The left panels display the archival data and \swift campaign, with colors going from dark red to dark blue, in chronological order. The right panels start from the last observation shown in the left panels, for easier comparison, and thus contain the last \swift observation and the \nustar campaign. Since that observation is colored in dark blue, the color order is opposite, and it goes from dark blue to yellow in chronological order. The thick lines connect consecutive observations, and go from the best-fit value of one observation into the best-fit value of the next one, chronologically. The legend shows the $\Delta t$ between the two connected observations.

Both \bor and \uxc show variability between the same sets of observations. The display chosen in Fig. \ref{fig:nh_contours_borus} and \ref{fig:nh_contours_uxc} is particularly useful to see in which timescales there is \nhlos variability. In chronological order, there is no variability in $\Delta t = 2$~days between the \xmm observations, given how the contours overlap. There is variability in the 3.8~yr timescale between the \xmm and \nustar observations, but again no variability in the shorter $\Delta t = 9$~days between the two archival \nustar observations. We note that these two \nustar  contours do not overlap, but there is a small \nhlos range that would overlap when projecting the contours onto the y axis. Following this logic, we can observe that there is variability in the 1.1~yr timescale, the 5.8~yr timescale (albeit very marginal; the \bor contours almost touch), the 14 day timescale between the first 2 \swift observations, and the 3 month timescale between the last two \swift observations. Moving into the right panels, there is variability in the 1.1~yr timescale between the last \swift observation and the \nustar campaign. As for the \nustar campaign itself, while there is variability between the first \nustar observation and the following 4 observations, with a minimum timescale of 18~days, the last 4 observations show no variability.

The variability between different observations can also be appreciated in the spectra shown in the Appendix (Figs. \ref{fig:spectra_borus} and \ref{fig:spectra_uxc}), as they are compared to the \chandra spectrum. It is easy to see that, even in the low-quality \swift observations, the slope in the $2-5$~keV range is markedly different between "\swift 1" and the \chandra spectrum, or "\swift 5" and the \chandra spectrum.

{In summary, Mrk 477 is observed to be \nhlos-variable over a large range of timescales: years, months, and as little as 14 days. With the data on Mrk 477 alone it is not possible to determine the most likely timescales of variability. However, Sect. \ref{nhlosvar_fullsample} shows that, when adding these data to the rest of the sample, \nhlos variability is more likely at longer timescales.}

\subsection{Probability of \nhlos variability vs. timescale}\label{nhlosvar_fullsample}

Mrk 477 has 15 different observations, which allows us to compare a large amount of epochs against each other. The total number of observation pairs is:

\begin{equation}
   _{15}C_{2} = \sum_{n=0}^{14} n = 105  .
\end{equation}
Out of these, using the values shown in Table \ref{tab:nhvar}, we calculate that 42 pairs are \nhlos-variable. That is, 42 comparisons of one epoch against another are incompatible with having the same \nhlos at a 90\% confidence level. This results in a probability of finding a variable pair, when comparing any two random observations, of $\sim40$\%. 

Therefore, even if we define Mrk 477 as an `\nhlos-variable source', the probability of finding it variable using only 2 observations is $<50$\%. However, the probability of drawing at least one pair of  \nhlos-variable observations grows rapidly with the number of observations taken, being 78\% when having 3 observations (3 comparisons) and 95\% when having 4 observations (4 comparisons).

\begin{figure*}
    \centering
    \hspace{-0.65cm}
    \includegraphics[scale=0.55]{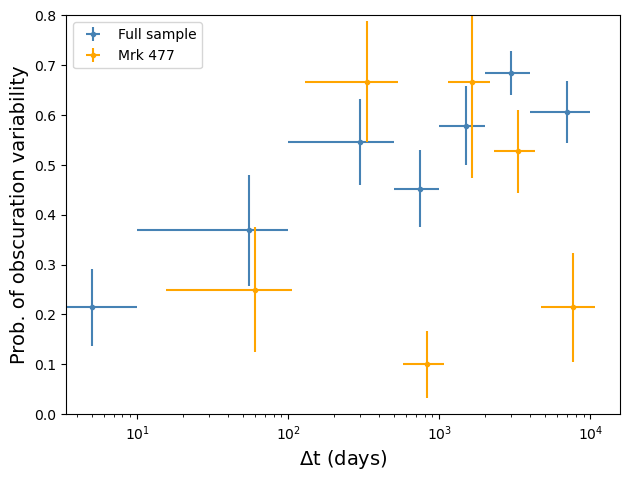}
    \includegraphics[scale=0.55]{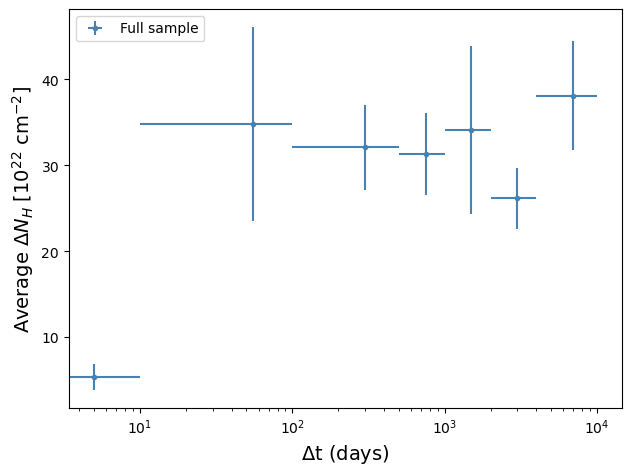}
    \caption{\textit{Left:} Fraction of observation pairs with incompatible \nhlos (at 90\% confidence) as a function of the time separation between observations. {The data shown in blue includes all sources within the parent sample analyzed so far \citep[27 sources, split into][ and this work]{Pizzetti2022,Torres-Alba2023,Pizzetti2024,Sengupta2024}. The data shown in orange includes only the observation pairs for Mrk 477}. \textit{Right:} Average change in \nhlos for all variable observation pairs as a function of the time separation between observations.}
    \label{fig:var_timescales}
\end{figure*}


With Mrk 477 alone it is not easy to determine whether this probability is affected by timescales (i.e. the time differences between compared observations). However, when adding the large amount of observation pairs for Mrk~477 to the rest of the sample \citep{Pizzetti2022,Torres-Alba2023,Pizzetti2024,Sengupta2024}, the total number of pairs available is $\sim350$. For the first time, we have sufficient statistics to determine the probability of \nhlos variability as a function of timescale.

We split the data in bins, such that each bin contains between 30 and 100 observation pairs. The uneven split is due to a much lower number of observations having smaller $\Delta t$, which would result in very poor time resolution if split evenly. Within each bin, and for all sources, we calculate the fraction of observation pairs which are incompatible with having the same \nhlos at a 90\% confidence level. The results are shown in Fig. \ref{fig:var_timescales}. There is a clear trend showing increased probability of \nhlos variability with larger timescales, increasing from $\sim20$\% at $\Delta t<10$~d to $\sim60-70$\% at $\Delta t>2000$~d.


Fig. \ref{fig:var_timescales} (right) also shows the average $\Delta$\nhlos for all observation pairs in the full sample that present \nhlos variability at a 90\% confidence level, as a function of timescale. The average $\Delta$\nhlos is between $30-40\times 10^{22}$ cm$^{-2}$ regardless of timescale, with the exception of very short timescales. For $\Delta t<10$~days, the average variability is much smaller ($5\times 10^{22}$ cm$^{-2}$), with very little variance across all observations. With only 6 variable pairs, this may be a result of poor statistics, and more data is likely needed to confirm this fact.

{Figure \ref{fig:var_timescales}, however, also shows the probability of finding a variable pair as a function of timescale for Mrk~477 only. As mentioned above, the data for this source contributes a sizable amount of the total observation pairs of the sample (almost a third). As such, it potentially introduces bias into the results. The Mrk 477 data has large spikes in the plot at timescales of $\sim2$~yr and $\sim 10$~yr, while going to very low probability at timescales of $\sim3$~yr. This is a result of the uneven sampling brought by the monitoring campaigns, with up to 5 observations taken very close together (twice) and then a few observations largely separated in time. Since the 5 observations grouped together tend to have much closer \nhlos, if one of the lone observations has a different \nhlos, this counts as 5 variable pairs at roughly the same timescale of separation. Therefore, our statistical approach is not valid when looking at a single source.}

{The effect the Mrk~477 has on the trends of the full sample can be perceived by eye, particularly at the timescales mentioned above. While it does not appear to dominate the trend of the full sample, it is worth being aware of its influence. This effect will be minimized as more data is added as a result of increasing the sample size.}

\subsection{\nhlos variability and BLR clouds}\label{blr_var}

{The origin of obscuration variability in AGN remains generally unknown, mostly due to the lack of constant monitoring in the X-ray regime. As such, the only transits that have a firm, associated origin are those that are short enough to be fully characterized within one single observation. All of these, which happened in timescales $<1$~d, have been firmly associated to BLR clouds \citep[e.g.][]{Elvis2004,Puccetti2007,Maiolino2010}.}

{Given how the only confirmed origin of obscuration variability is the BLR, one cannot rule out that it is responsible for all obscuration variability, independently of timescale. That is not to say that a BLR cloud transit lasts for a $\Delta t$ of years, but rather that two independent observations taken a given $\Delta t$ apart have intercepted completely different BLR clouds. On the other hand, long-timescale variability could simply originate in larger structures (i.e. the torus), that would have longer transit times, even for a cloud/clump of the same size as a BLR cloud, due to lower velocities.} 

{In this section, we explore how easily BLR clouds could explain the behavior observed in Fig. \ref{fig:var_timescales} (if at all) in order to potentially rule them out as responsible for obscuration variability at all timescales.}

{For starters, the fact that there is a dependence of the probability of \nhlos variability with timescale already suggests that BLR clouds may not be the exclusive origin of such variability.} This is because, above $\Delta t$ larger than a single cloud eclipsing event {(i.e. the cloud crossing time, $t_{\rm cross}$)}, the probability of observing a difference in \nhlos should be independent of timescale\footnote{Provided the cloud distribution is homogeneous.}. As BLR clouds orbit the SMBH at very close distances, one expects a very quick transit in and out of the line-of-sight \citep[e.g., $t_{\rm cross}\sim4$~ks $\simeq1$~h, as observed by][]{Maiolino2010}. This crossing time is much smaller than the timescales of variability discussed here, and thus one would always observe a different BLR cloud and obtain a completely unrelated \nhlos measurement. Therefore, the probability of measuring \nhlos variability should be independent of timescale at $\Delta t \gg 1$~h. {That is, for all bins in Fig. \ref{fig:var_timescales}, the probability should not be dependent on timescale, and instead remain constant. As this is not the case, this figure favors the origin of obscuration variability being at torus scales.}

{However, in order to be conservative, we explore the possibility that $t_{\rm cross}$ could actually be much longer than 1~h, and perhaps the lack of such detections is because current observations are biased toward detecting quick transits (i.e. only those that can occur within one single observation, which are at most hundred of ks, which is a few days).}

{As already stated above, the probability of detecting variability should remain constant at $\Delta t > t_{\rm cross}$ (if all variability originates in the BLR). Looking at Fig. \ref{fig:var_timescales}, one could consider that the increase of probability with timescale} holds up to about $\Delta t \sim 300-500$~d, and  that {for larger timescales the probability is compatible with being roughly constant (at an average rate of $\sim60-70$\%, within errors)}. In such a scenario, and as described above, the timescale of $\sim300-500$~d would correspond to the typical crossing time of an individual BLR cloud. If one assumes Keplerian, circular orbits, and a reasonable SMBH mass range (M$_{\rm SMBH}=10^{7}-10^{8}$~M$_\odot$), it is possible to obtain a rough approximation of the BLR cloud sizes by simply doing:
\begin{equation}
    v = \sqrt{\frac{GM_\mathrm{SMBH}}{r}},
\end{equation}
where $v$ is the Keplerian velocity, G is the gravitational constant, and r is the 
distance to the SMBH. The cloud size ($R_{\rm c}$) is then computed as $v=R_{\rm c}\times t_{\rm cross}$.

\begin{figure}
    \centering
    \hspace{-0.65cm}
    \includegraphics[scale=0.55]{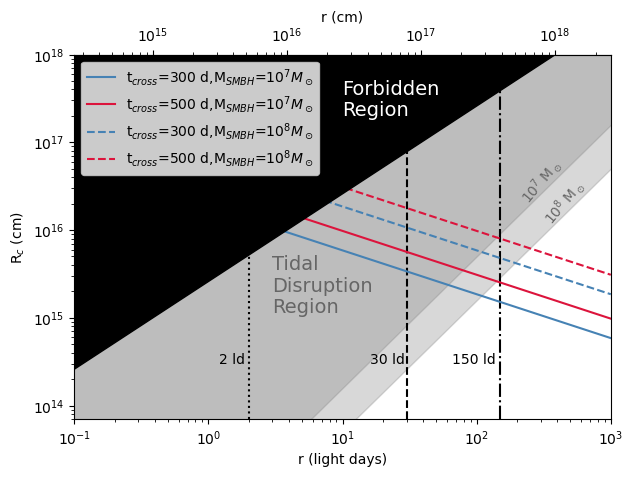}
    \caption{BLR cloud sizes as a function of radial distance, {computed under the assumption of two different M$_{\rm SMBH}$ and a cloud crossing time of} $t_{\rm cross }=$300 d (blue) and $t_{\rm cross }=$500 d (red). {The black shaded area corresponds to a `forbidden region', an area removed from the plot as the clouds in it would be larger than their distance to the SMBH. The gray shaded areas correspond to `tidal disruption regions', where clouds would not remain stable due to the gravitational influence of the SMBH. The vertical lines correspond to the minimum, average, and maximum BLR sizes observed by the GRAVITY collaboration \citep{Gravity2023}.}}
    \label{fig:BLRsizes}
\end{figure}

{We also consider that such large clouds cannot exist stably arbitrarily close to the SMBH, since its gravitational influence would result in tidal disruption. We use Eq. 5 in \cite{Ichikawa2017} in order to calculate the radius below which clouds would be unstable:}
\begin{equation}
    R_{\rm c} = 4.8\times10^{-3} \left( \frac{c_s}{3 \ \mathrm{km  \ s^{-1}}} \right) \left( \frac{r}{1 \ \mathrm{pc}} \right)^{3/2} \left( \frac{M_{\rm SMBH}}{10^8 \ \mathrm{M_\odot}} \right)^{-1/2}, 
\end{equation}
{in units of parsec}, where $c_s$ is the speed of sound in the medium, computed as $c_s = \sqrt{k_BT/\mu m_p}$. We assume a BLR cloud temperature of roughly $\sim 10^4$~K \citep[see e.g.][]{Ilic2008}, a mean molecular weight $\mu=0.5$ as is suitable for the ionized gas of the BLR \citep{Muller2022}, $k_B$ is the Boltzmann constant {and $m_p$ is the proton mass.}

Fig. \ref{fig:BLRsizes} shows the results of this simple calculation for all reasonable cloud distances to the SMBH. The vertical lines show BLR sizes as estimated by the GRAVITY collaboration for 26 AGN \citep[smallest size R$_{\rm BLR}=2$~ld, largest size R$_{\rm BLR}=150$~ld, and average BLR size R$_{\rm BLR}\sim30$~ld,][]{Gravity2023}. In all instances, the BLR cloud sizes remain $\sim2-3$ orders of magnitude larger than previous estimates \citep[$\sim10^{13}$~cm, e.g.,][]{Maiolino2010,Netzer2015}. {Moreover, the majority of the area in the plot falls within a region within which the clouds would likely be tidally disrupted. Only for black hole masses $M_{\rm SMBH}\leq 10^7 \ M_{\odot}$, and if the BLR has an above-average size, clouds this large can stably exist.}

{The other obvious possible explanation is that BLR clouds are only responsible for variability at much shorter timescales than $300-500$~d, after which the torus takes over as the main origin. This hypothesis is supported by Fig. \ref{fig:var_timescales} ({Right}),  which shows a different average $\Delta$\nhlos only at $\Delta t <10$~d. That is, it could be that obscuration variability at $\Delta t<10$~d is the only one caused exclusively by BLR clouds}, and that their average column density is about $5\times 10^{22}$ cm$^{-2}$. This would align with the results of \citet{Maiolino2010}, who observed $\Delta$\nhlos$=5-10\times 10^{22}$~cm$^{-2}$ in the two consecutive eclipses attributed to BLR clouds. {Indeed}, repeating the calculations shown in Fig. \ref{fig:BLRsizes} with $t_{\rm cross}=10$~d would lead to a cloud size of $\sim 5\times10^{13}$~cm at 150~ld, which, again, is much more compatible with previous observations.

However, {it is worth mentioning that,} using this simplified method, we cannot rule out that a complex cloud distribution (i.e. with cloud density/size depending on the distance to the SMBH) could reproduce the variability pattern in Fig. \ref{fig:var_timescales}. In future works, we will explore the exact BLR properties that could reproduce it {(if any)}. 

\subsection{Variability Results in the Context of Previous Work}\label{other_works}





{The work presented here uses one of the largest samples to date, if not the largest, when it comes to exploring the typical timescales and values of obscuration variability. As such, comparing the results shown in Fig. \ref{fig:var_timescales} to previous works is not trivial. Indeed, most works on obscuration variability to date focus on single sources \citep[e.g.][]{Pizzetti2022,Kayal2023}. Furthermore, they also tend to focus on extreme sources rather than the bulk of the population; those known as `Changing-look AGN', which transition from obscured to unobscured  \citep[Seyfert 1 to Seyfert 2; see e.g.,][for a review]{Titarchuk2024}, or from Compton-thin to Compton-thick \citep[across the \nhlos$=10^{24}$~cm$^{-2}$ threshold, e.g.,][]{Risaliti2005,Bianchi2009,Guainazzi2012,Marchese2012,Miniutti2014,Ricci2016,Marchesi2022,Serafinelli2023}}.

{As such, the sample selection is markedly different between the results presented here and those obtained in most previous works. In fact, based on the cited works\footnote{This includes comparison between all observation pairs that are incompatible in \nhlos, within errors in \cite{Bianchi2009,Marchese2012,Miniutti2014,Ricci2016,Marchesi2022}. Other works did not present a comprehensive list of \nhlos values for all observations of the source, and where thus excluded from this calculation.}, the average column density variability in Changing-look AGN is $\langle$\nhlos$\rangle\sim 7\times10^{23}$~cm$^{-2}$. This is more than twice the average found for our sample, which is not surprising, given that the mentioned works specifically select sources that have large obscuration variability. Similarly, \cite{Laha2020} studied a sample of $\sim20$ sources, most of which had an average column density of $1-10\times10^{22}$~cm$^{-2}$. The average column density variability of the analysis perfomed in their work is $\langle$\nhlos$\rangle < 5\times10^{22}$~cm$^{-2}$.}

{These comparisons point to the fact that $\Delta$\nhlos is likely dependent on \nhlos, and somewhat proportional to it. Therefore, our average obscuration column density variability of $\sim3\times10^{23}$~cm$^{-2}$ is valid for sources with typical \nhlos values in the $\sim10^{23}-10^{24}$~cm$^{-2}$ range.}

{Finally, our results showing that the probability of obscuration variability increases with time are in agreement with those of \cite{Markowitz2014}. \cite{Markowitz2014} observed a total of 8 sources for 17~yrs, detecting 8 individual eclipsing events. While this is a low number, it remains the largest sample to date with fully-observed eclipsing events (i.e. ingress to egress). For timescales larger than 10 days, an increase in the probability of eclipses is observed for both type I and type II AGN. Our results, which do not observe full eclipsing events but are based on a much larger sample of sources and number of events, significantly strengthen their findings. }

\section{Conclusions}\label{conclusions}

In this work, we have analyzed 15 X-ray observations of Mrk 477, taken with four different telescopes, spanning a total of $\sim13$~yrs, and with a large variety of time differences between consecutive observations. In here, we summarize our main conclusions:
\begin{itemize}
    \item Mrk 477 presents \nhlos variability, between multiple sets of observations, and with timescales as short as $\sim2$~weeks and as long as years.
    \item Although we define Mrk 477 as an `\nhlos-variable source', the probability of finding \nhlos variability between any two randomly drawn observations is $\sim40$\%. We caution that this rough estimate does not take into account the timescales between different observations, and that more data would be needed to give a statistically meaningful number as a function of $\Delta t$.
    \item When adding the results of this work to those of another 26 sources analyzed by ourselves, we conclude that \nhlos variability is more likely to occur at yearly timescales than at shorter ones ($\sim$days/weeks). The probability of observing a variable pair increases from $\sim20$\% at $\Delta t<10$~days, to $\sim60-70$\% at timescales larger than 5 years.
    \item The average $\Delta$\nhlos between variable observations is $30-40\times 10^{22}$~cm$^{-2}$ regardless of timescale, with the exception of $\Delta t<10$~days, for which we find an average variation of $5\times 10^{22}$~cm$^{-2}$.
    \item Mrk 477 is best described with a thin reflector with large covering factor by \bor. \uxc qualitatively agrees with this description, given how it does not requires a Compton-thick reflector, and \sigtor is at the highest possible value.
    \item We test the hypothesis of BLR clouds resulting in all of the observed \nhlos variability, for all sources. If we assume a homogeneous cloud distribution, we can easily attribute \nhlos variability to BLR clouds at timescales $\Delta t<10$~d. On the other hand, under the same hypothesis, BLR clouds would need to have sizes $>10^{15}$~cm to explain variability at larger timescales. This is two orders of magnitude larger than previously observed/assumed. However, we cannot rule out that a particular distribution of cloud sizes and density as a function of radial distance to the SMBH could reproduce the data.  
    \item Mrk 477 shows no variability in reflection parameters (that is, geometrical parameters of the torus, such as \cf, \nhav, \sigtor) when splitting the full observation dataset into two sub-sets. The same is true for the photon index. 
\end{itemize}

\vspace{1cm}
\acknowledgments

We thank the referee for their useful comments and suggestions, and their careful reading of the draft. N.T.A., M.A., and I.C. acknowledge funding from NASA under contracts 80NSSC23K1611, 80NSSC23K0484, and 80NSSC23K0484. The scientific results reported in this article are based on observations made by the X-ray observatories \nustar, \xmm, \chandra and \swift, and has made use of the NASA/IPAC Extragalactic Database (NED), which is operated by the Jet Propulsion Laboratory, California Institute of Technology under contract with NASA. We acknowledge the use of the software packages XMM-SAS and HEASoft.

\vspace{2cm}
\bibliographystyle{aasjournal}
\bibliography{bibliography}

\newpage
\appendix

\section{Individual spectra}\label{App:spectra}

In this section we present spectra for each individual observation, fit with both \bor and \uxc, in Figs. \ref{fig:spectra_borus} and \ref{fig:spectra_uxc}, respectively. All spectra were fit simultaneously, but we opt to show them separately in order to avoid overcrowding the plot. For easier comparison, and in order to appreciate the time variability, all spectra are compared to the same \chandra best-fit, as shown in Fig. \ref{fig:chandra_spectra}. The colors used to plot the spectra match, observation-wise, the contours in Figs. \ref{fig:nh_contours_borus} and \ref{fig:nh_contours_uxc}. The spectra are displayed in chronological order.

\begin{figure*}
    \centering
    \hspace{-0.65cm}
    \includegraphics[scale=0.47]{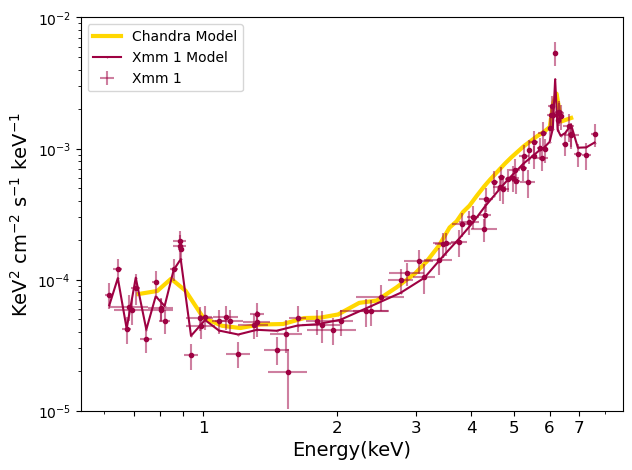}
    \includegraphics[scale=0.47]{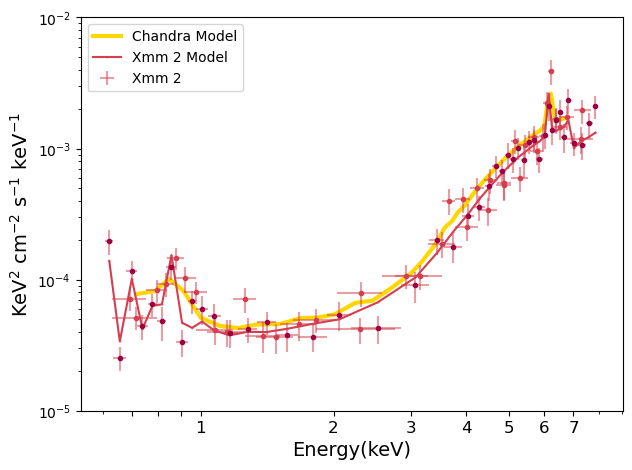}
    \includegraphics[scale=0.47]{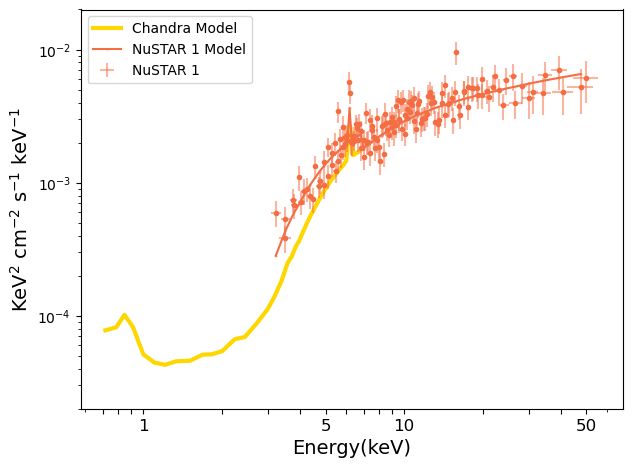}
    \includegraphics[scale=0.47]{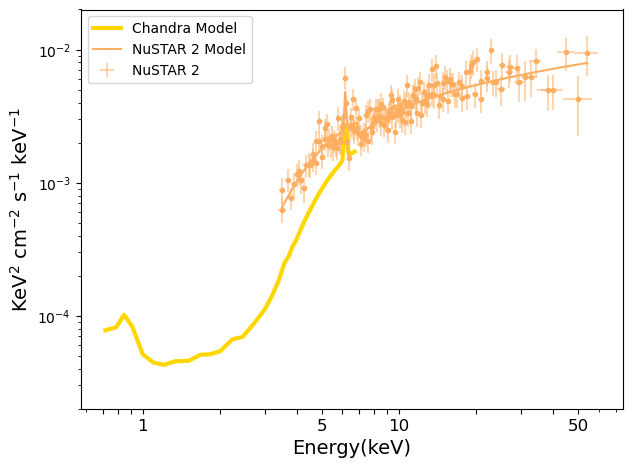}
    \includegraphics[scale=0.47]{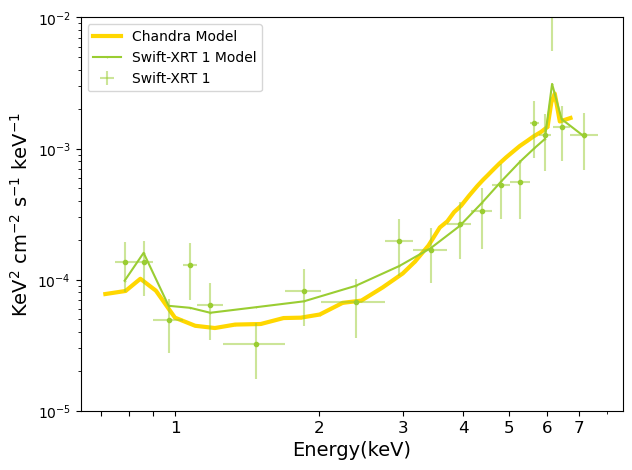}
    \includegraphics[scale=0.47]{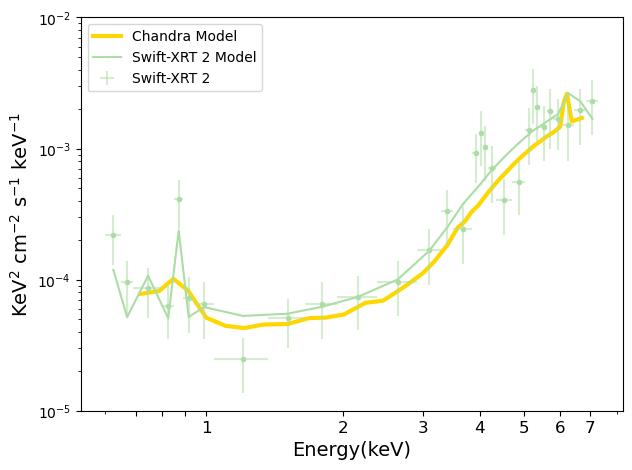}
    \includegraphics[scale=0.47]{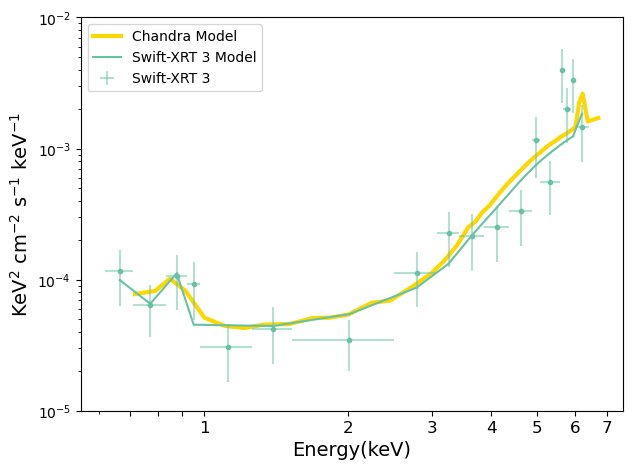}
    \includegraphics[scale=0.47]{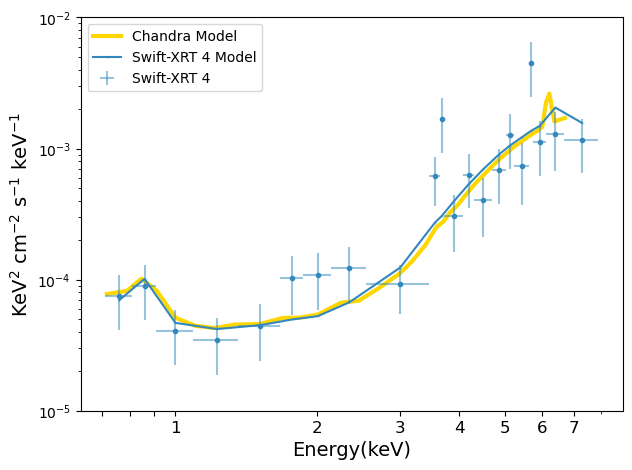}
    \caption{All observations used in this work, except for the \chandra observation already shown in Fig. \ref{fig:chandra_spectra}, as fit with \bor. They are individually compared to the reference \chandra observation. The spectra are shown in chronological order.}
    \label{fig:spectra_borus}
\end{figure*}

\renewcommand{\thefigure}{\arabic{figure} (Cont.)}
\addtocounter{figure}{-1}

\begin{figure*}
    \centering
    \hspace{-0.65cm}
    \includegraphics[scale=0.47]{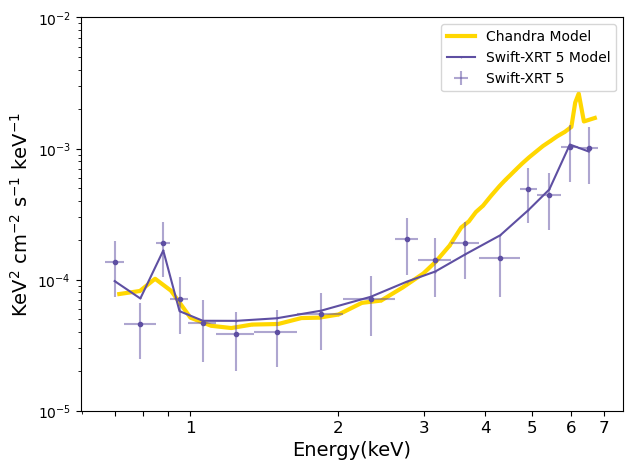}
    \includegraphics[scale=0.47]{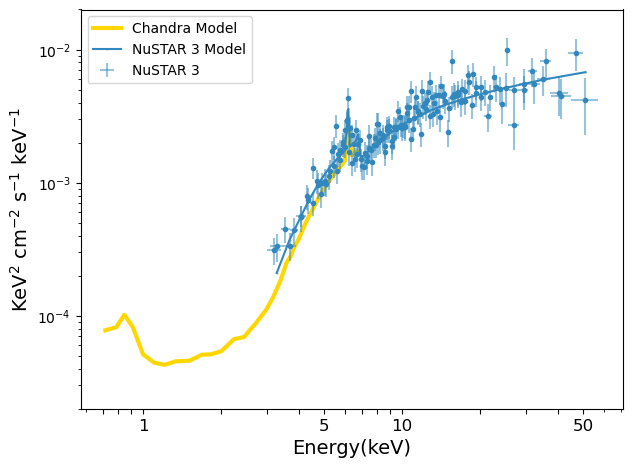}
    \includegraphics[scale=0.47]{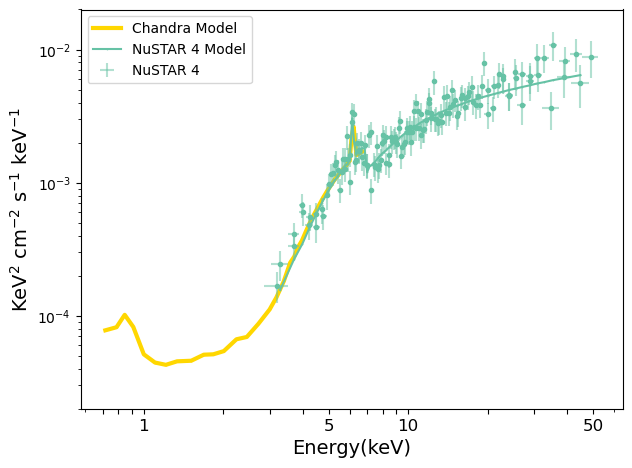}
    \includegraphics[scale=0.47]{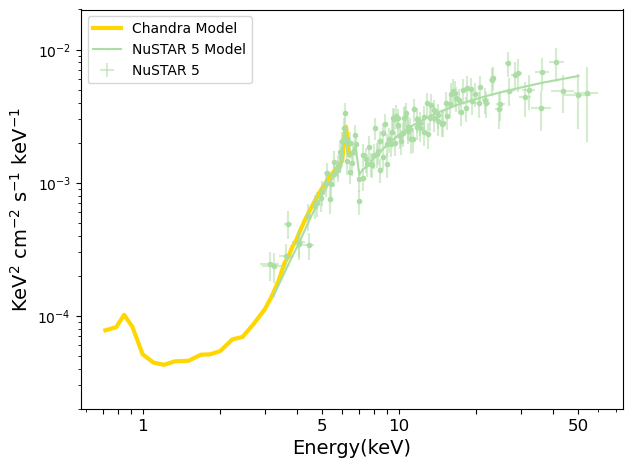}
    \includegraphics[scale=0.47]{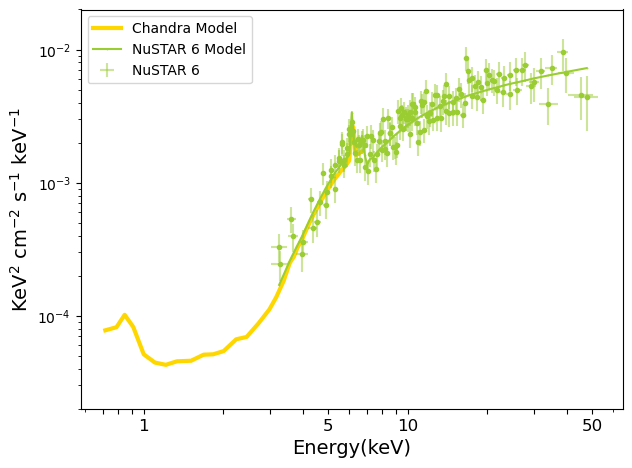}
    \includegraphics[scale=0.47]{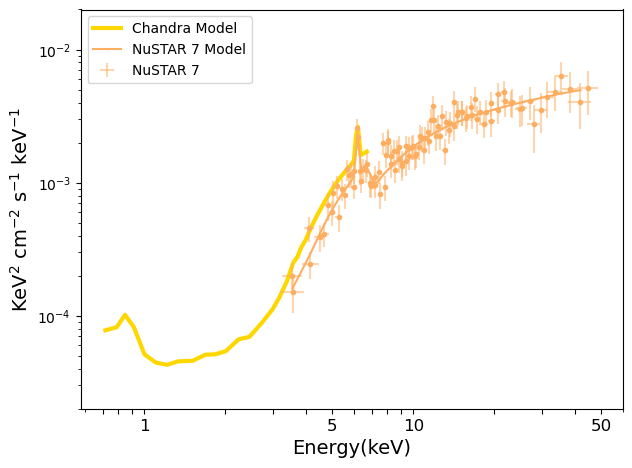}
    \caption{}
\end{figure*}

\renewcommand{\thefigure}{\arabic{figure}}

\begin{figure*}
    \centering
    \hspace{-0.65cm}
    \includegraphics[scale=0.47]{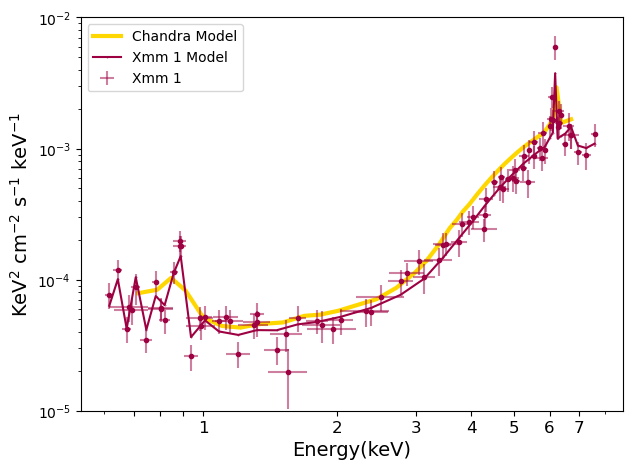}
    \includegraphics[scale=0.47]{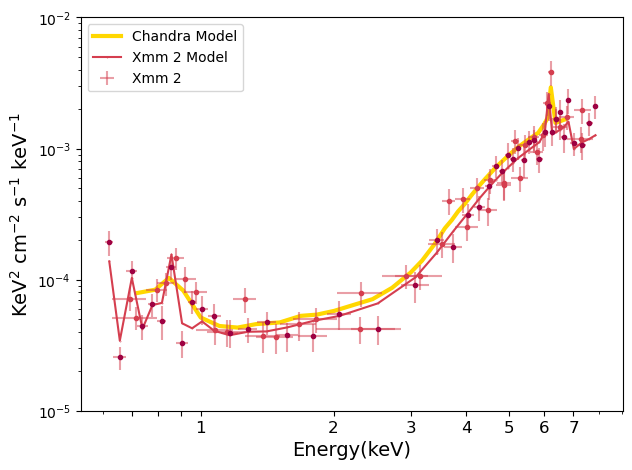}
    \includegraphics[scale=0.47]{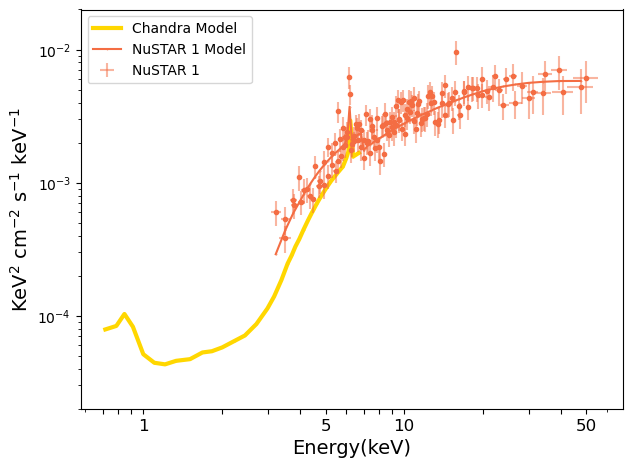}
    \includegraphics[scale=0.47]{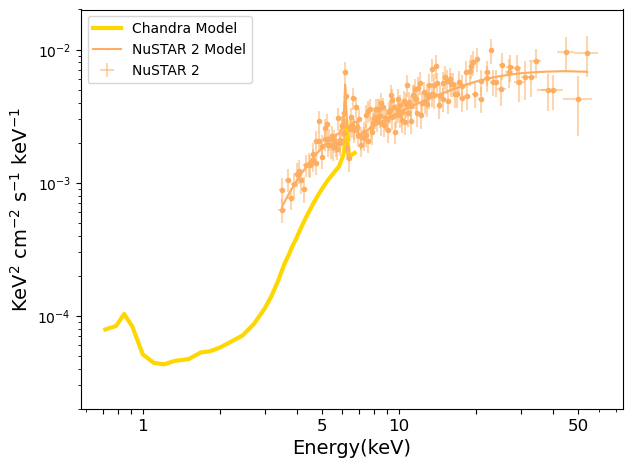}
    \includegraphics[scale=0.47]{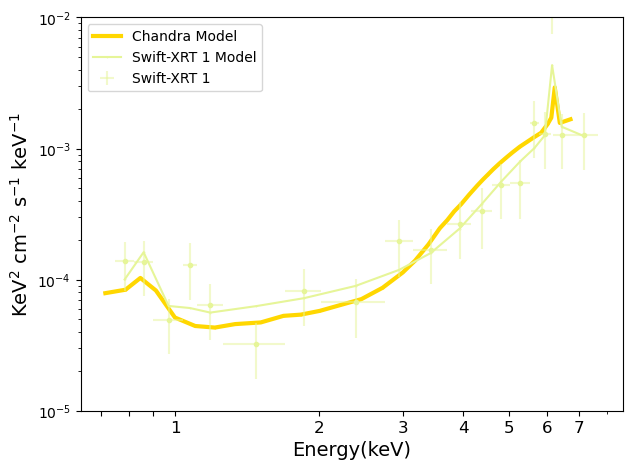}
    \includegraphics[scale=0.47]{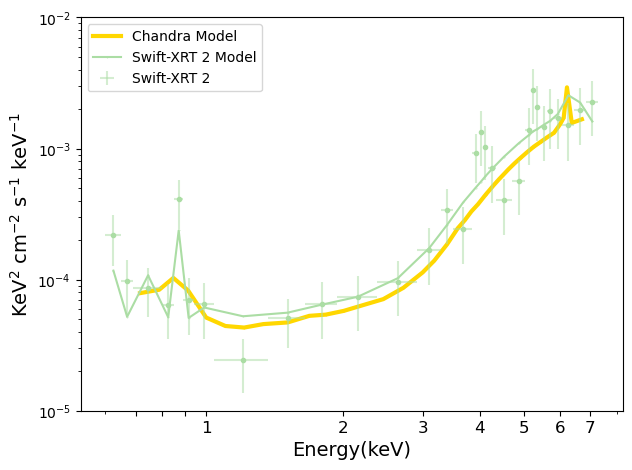}
    \includegraphics[scale=0.47]{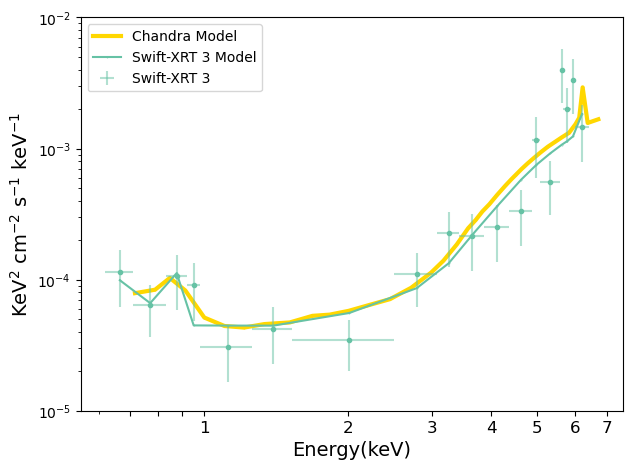}
    \includegraphics[scale=0.47]{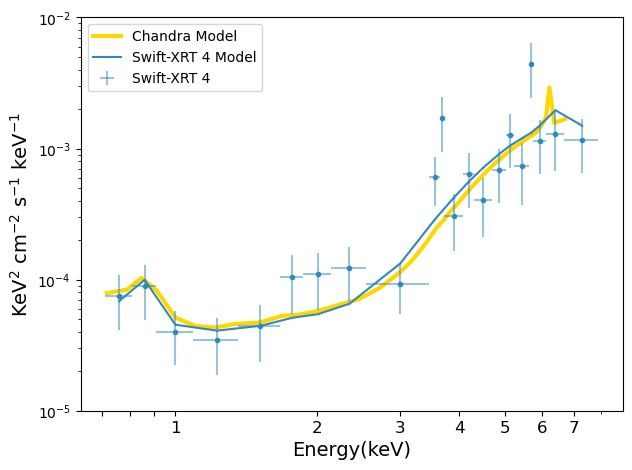}
    \caption{All observations used in this work, except for the \chandra observation already shown in Fig. \ref{fig:chandra_spectra}, as fit with \uxc. They are individually compared to the reference \chandra observation. The spectra are shown in chronological order.}
    \label{fig:spectra_uxc}
\end{figure*}

\renewcommand{\thefigure}{\arabic{figure} (Cont.)}
\addtocounter{figure}{-1}

\begin{figure*}
    \centering
    \hspace{-0.65cm}
    \includegraphics[scale=0.47]{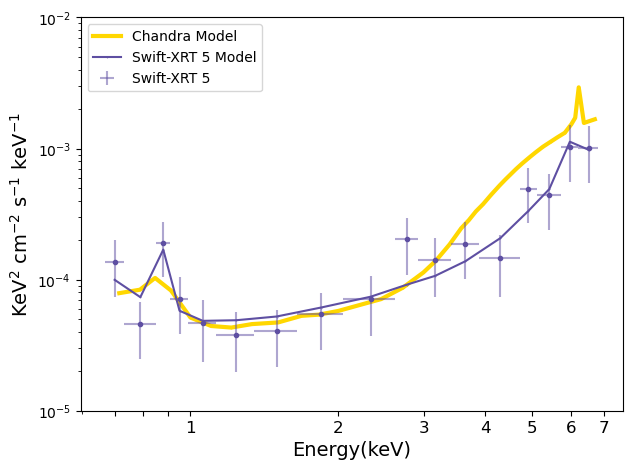}
    \includegraphics[scale=0.47]{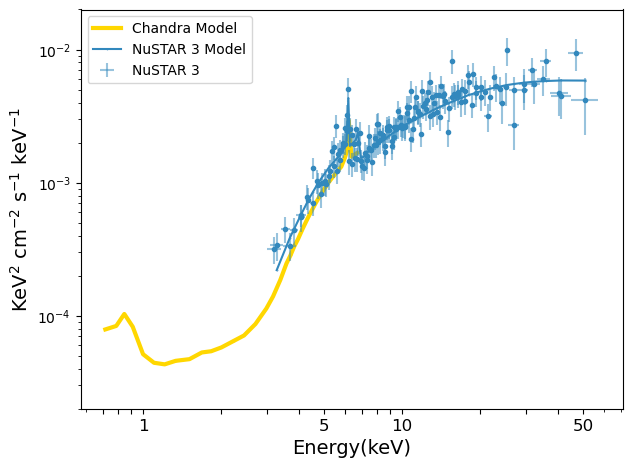}
    \includegraphics[scale=0.47]{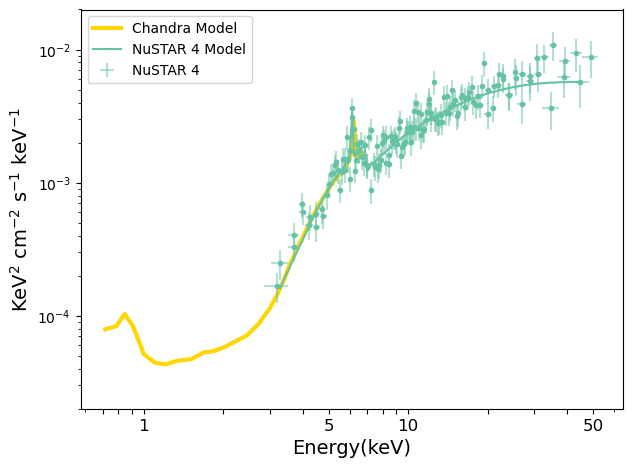}
    \includegraphics[scale=0.47]{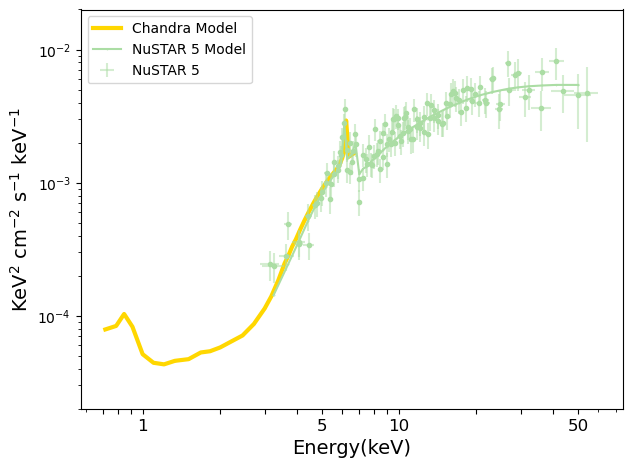}
    \includegraphics[scale=0.47]{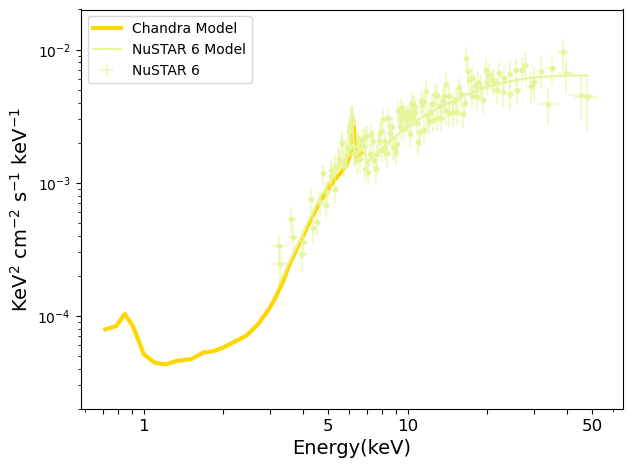}
    \includegraphics[scale=0.47]{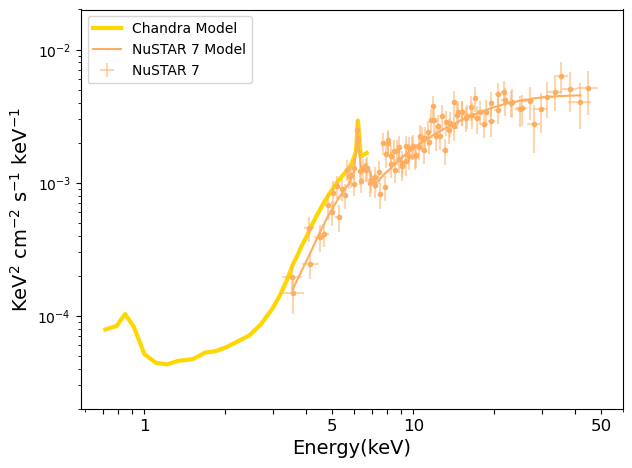}
    \caption{}
\end{figure*}

\renewcommand{\thefigure}{\arabic{figure}}

\section{Non-Variability of the Torus Geometry}\label{App:reflection}

One of the fundamental assumptions of the fitting process described in this work, as well as in the analysis of the parent sample in \cite{Pizzetti2022,Torres-Alba2023,Pizzetti2024}, is that one can consider that the reflection parameters do not change in timescales of $\sim10-20$~yrs. This is a necessary assumption when dealing with monitoring campaigns that do not have simultaneous soft ($<10$~keV) and hard ($>10$~keV) coverage, or with archival data\footnote{This is the vast majority of observations for AGN in X-rays, with only about $\sim10$\% of \nustar time having simultaneous observations with either \chandra, \swift or \xmm (see Fig. 6 in \citealt{Boorman2024}).}. Indeed, the full band is needed to constrain parameter degeneracies and separate \nhav variability from intrinsic luminosity variability\footnote{And even under the mentioned assumption, it is sometimes impossible to disentangle between the two, see \cite{Torres-Alba2023,Pizzetti2024}.}. 

With the modeling described in this work, as well as in the aforementioned works, it is possible to break the degeneracy because full-band coverage is available, even if not simultaneous. That is, under the assumption of non-variability of the reflector. However, a few bright and well-known sources are known to show reflection variability in similar timescales \citep[see e.g.][and references therein]{Boorman2024}. It is currently unknown how common this phenomenon is, although it is certain that it would require the reflector to have a scale much smaller than the scales of the torus ($\ll1-10$~pc) to greatly vary in the observed short timescales. The current, most common assumption in the literature is that reflection occurs all throughout the torus and, in such a case, the non-variability assumption is valid.

This is not easy to test for the large majority of sources in the archive, given how good quality \nustar observations are needed in multiple, different epochs \cite[e.g.][]{Boorman2024}. Furthermore, a large number of observations at each epoch (or great data quality) may also be needed in order to break degeneracies and reduce errors of the derived parameters \citep[see e.g.][]{Marchesi2022}, to allow for a comparison.

In this work, we are putting together archival observations as well as two different monitoring campaigns. Thanks to the large amount of data, we can test the validity of this hypothesis for Mrk 477, and for this specific dataset. Due to data quality and limited band coverage, the \swift campaign cannot be used to constrain reflection parameters on its own. Therefore, we have split the analysis into two datasets that have a similar total number of counts: the archival data + the \swift campaign, and the \nustar campaign alone. Both sets contain \nustar observations, which are key to estimating reflection parameters. 

Table \ref{tab:reflection} contains the results for the parameters considered to be common among all observations for three different fits: the two separate datasets mentioned above, and the full dataset. For the purposes of this comparison, we only show the common parameters in Table \ref{tab:reflection}, but the result of the full fit (for the split datasets) can be found in the Appendix (see Tables \ref{table:split_archival} and \ref{table:split_nustar}). The determinations for the parameters that vary in each observation, for the full dataset fit, can be found in Sect. \ref{obscuration_variability}.

Figures \ref{fig:reflection_contours_uxc} and \ref{fig:reflection_contours_borus} show contour plots at {90\% and 99\% confidence level} of the photon index $\Gamma$ against each of the reflection parameters, for \uxc and \bor, respectively. Both the table and figures show remarkable agreement between the parameter contours of the two datasets, and of the full dataset.

\begin{figure*}
    \centering
    \hspace{-0.65cm}
    \includegraphics[scale=0.5]{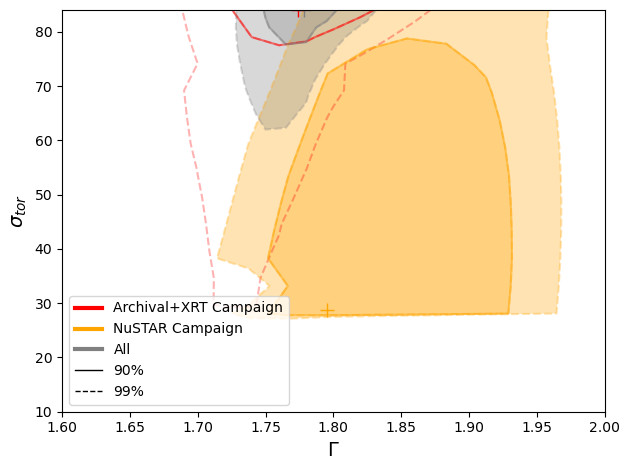}
    \includegraphics[scale=0.5]{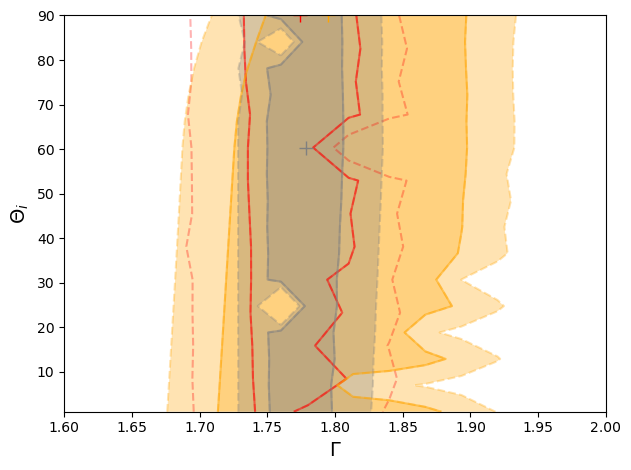}
    \caption{Confidence contours (at 1$\sigma$, 2$\sigma$ and 3$\sigma$) for the width of the torus cloud distribution (left), and inclination angle (right), against the photon index, for the \uxc model. The different colors represent the different datasets used to test the hypothesis that the reflection parameters do not change: the archival data + \swift campaign (red), the \nustar campaign (blue), and the full dataset (green). The colored crosses indicate the best fit. The three datasets are in good agreement, with the parameter determination becoming more tightly constrained when using the full dataset.}
    \label{fig:reflection_contours_uxc}
\end{figure*}

\begin{figure*}
    \centering
    \hspace{-0.65cm}
    \includegraphics[scale=0.5]{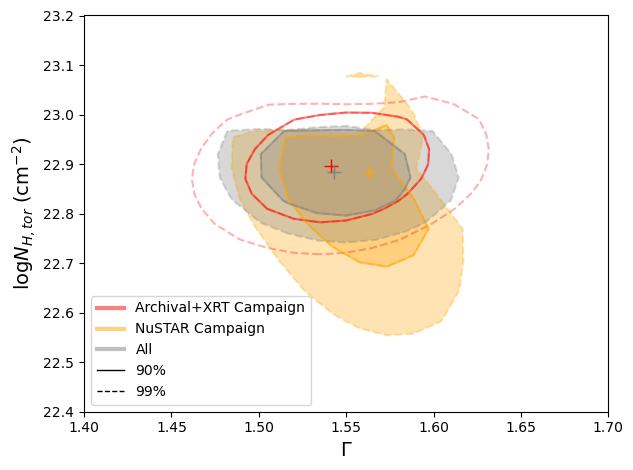}
    \includegraphics[scale=0.5]{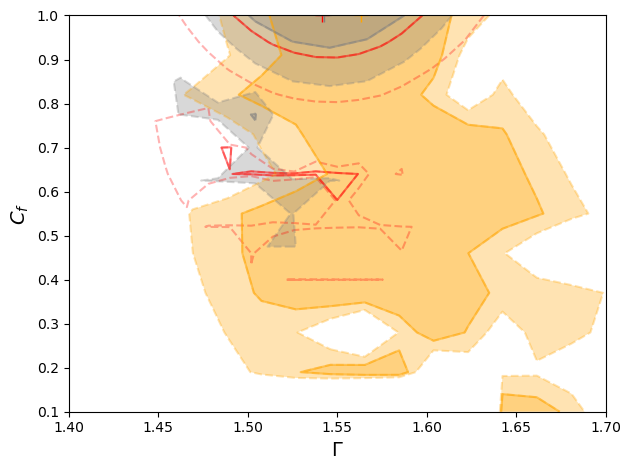}
    \includegraphics[scale=0.5]{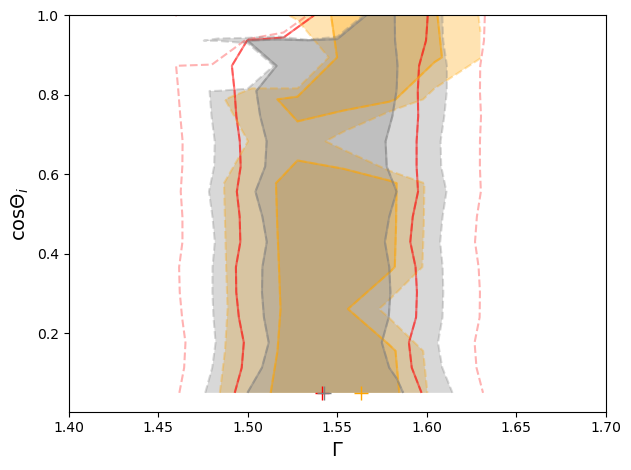}
    \caption{Confidence contours (at 1$\sigma$, 2$\sigma$ and 3$\sigma$) for the width of the cloud distribution (top left), covering factor (top right), and inclination angle (bottom), against the photon index, for the \bor model. The different colors represent the different datasets used to test the hypothesis that the reflection parameters do not change: the archival data + \swift campaign (red), the \nustar campaign (blue), and the full dataset (green). The colored crosses indicate the best fit. The three datasets are in good agreement, with the parameter determination becoming more tightly constrained when using the full dataset.}
    \label{fig:reflection_contours_borus}
\end{figure*}

For the \uxc model, the parameter that dictates the spread of the cloud distribution (\sigtor) has a different `best-fit' value in the \nustar-only campaign fit, with large errors. This is likely due to the fact that the reflector in Mrk 477 is subdominant, and thus difficult to constrain. However, the \nustar-only campaign agrees with the other two datasets {within the 99\% confidence level}. It can also be seen how the increase in number of observations results in much tighter constraints, favouring a higher value of \sigtor, close to the upper limit of \sigtor$=84$. The inclination angle is completely unconstrained, which is not uncommon, even for large datasets with multiple observations \citep[see e.g.][]{Torres-Alba2021,Pizzetti2024}. The contours become narrower for the full dataset, which is only a reflection of the tighter constraints on the photon index ($\Gamma$).

For the \bor model we see similar results, with the inclination angle being once more completely unconstrained. The determinations for \nhav are in very good agreement, with a small tightening of the contours as the whole dataset is used. The contours for the covering factor show a good agreement around the best-fit value (\cf$=1$), while also showcasing a complex parameter space, particularly for the \nustar-only campaign. As more data is added, these other areas of low $\Delta\chi^2$ are reduced, even if not completely eliminated.

Mrk 477 does not have a dominant reflector (as indicated by the very low \nhav of the \bor model, and as shown in Fig. \ref{fig:chandra_spectra}). This fact makes it additionally difficult to constrain some of the reflection parameters despite the abundance of data. 

Overall, the agreement between datasets is good. Meaning, the data is compatible with the reflection parameters not varying across observations. It is worth noting that, since the reflection parameters are mostly fit through the \nustar observations, the test performed compares the torus parameters in 2014 (date of the archival \nustar data) versus 2022 (date of the monitoring campaign by \nustar). Therefore, the reflection parameters are consistent in a timescale of about a decade. 

However, this is not to be taken as definitive proof that the reflector of Mrk 477 is non-variable, or further extrapolating that, since it is not variable, the reflection originates in the whole torus. Rather, this just serves as a consistency test of our assumed hypothesis (i.e. that we can impose non-variability of the reflection parameters for this particular work). In order to fully test this assumption and its physical implications, we would ideally need monitoring campaigns spread over different years, providing high-quality data in both soft and hard band, such that it is possible to tightly constrain the reflection parameters in each epoch. Further testing would also require a sizable sample of sources, to determine how common reflection variability is over non-variability. Sources with a higher reflection dominance would also serve as a better testing benchmark, compared to Mrk~477.

\section{Model Tables}\label{App:Tables}

Table \ref{tab:reflection} shows a comparison between the joint fit of all the data, and the joint fit of two sub-sets, in terms of the parameters kept constants at all epochs. In this section, we present tables including all of the parameters (kept constant and variable) in the two sub-sets used for the mentioned comparison: Table \ref{table:split_archival} shows the full fit for the archival plus \swift campaign; while Table \ref{table:split_nustar} shows the same for the \nustar-only dataset. The information added is thus equivalent to that shown in Table \ref{tab:nhvar} for the full dataset.

\begin{deluxetable}{c|c|c}
\label{table:split_archival}
\tablecaption{Best fit parameters for the archival data and \swift campaign of Mrk 477.}
\tablehead{\colhead{Model} & \colhead{\bor} & \colhead{\uxc} \\  }
\startdata
red $\chi^2$ & 1.04  & 1.06\\
$\chi^2$/d.o.f. & 622.5/596  & 629.9/596 \\
kT  & 0.29$^{+0.06}_{-0.05}$ & 0.30$^{+0.06}_{-0.05}$ \\
E$_{\rm line}$ & $^{+0.01}_{-0.01}$ & 0.91$^{+0.01}_{-0.01}$ \\
$\Gamma$  & 1.54$^{+0.05}_{-0.05}$ &  1.77$^{+0.06}_{-0.05}$ \\
$N_{H,av}$  & 0.08$^{+0.02}_{-0.02}$ & $-$ \\
\cf & 1.00$^{+u}_{-0.12}$ & $-$\\
\ctk  & $-$ & 0* \\
$\sigma_{\rm tor}$ & $-$  &  84.0$^{+u}_{-15.6}$ \\
Cos ($\theta_{Obs}$) & 0.05$^{+u}_{-u}$  & 0.00$^{+u}_{-u}$\\ 
F$_s$ (10$^{-2}$) & 2.46$^{+1.23}_{-0.46}$ & 18.9$^{+2.9}_{-3.2}$ \\
norm (10$^{-3}$) & 1.51$^{+0.37}_{-0.54}$ & 2.38$^{+0.50}_{-0.51}$ \\
\hline 
$N_{H,Ch}$  & 0.38$^{+0.05}_{-0.05}$  & 0.30$^{+0.05}_{-0.07}$  \\
$N_{H,xmm1}$  & 0.44$^{+0.04}_{-0.04}$  & 0.37$^{+0.04}_{-0.04}$  \\
$N_{H,xmm2}$  & 0.40$^{+0.03}_{-0.03}$  & 0.33$^{+0.03}_{-0.04}$  \\ 
$N_{H,nus1}$ & 0.21$^{+0.03}_{-0.03}$ & 0.19$^{+0.04}_{-0.03}$ \\
$N_{H,nus2}$  & 0.17$^{+0.02}_{-0.02}$  & 0.16$^{+0.03}_{-0.03}$ \\
$N_{H,swift1}$  & 0.60$^{+0.19}_{-0.13}$  & 0.53$^{+0.24}_{-0.17}$\\
$N_{H,swift2}$  & 0.33$^{+0.08}_{-0.06}$  & 0.27$^{+0.09}_{-0.06}$\\
$N_{H,swift3}$  & 0.41$^{+0.12}_{-0.10}$  & 0.35$^{+0.16}_{-0.11}$\\
$N_{H,swift4}$ & 0.33$^{+0.12}_{-0.08}$  & 0.25$^{+0.12}_{-0.08}$\\
$N_{H,swift5}$  & 0.82$^{+0.55}_{-0.20}$  & 0.75$^{+0.79}_{-0.25}$ \\\hline
$C_{Ch}$ & 1* & 1*  \\
$C_{xmm1}$  & 0.89$^{+0.14}_{-0.11}$  & 0.90$^{+0.17}_{-0.11}$  \\
$C_{xmm2}$ & 0.88$^{+0.14}_{-0.11}$  & 0.91$^{+0.16}_{-0.11}$  \\ 
$C_{nus1}$  & 0.90$^{+0.21}_{-0.14}$  & 1.08$^{+0.17}_{-0.11}$  \\
$C_{nus2}$  & 1.02$^{+0.23}_{-0.16}$ &  1.26$^{+0.31}_{-0.19}$ \\
$C_{swift1}$  & 1.31$^{+0.40}_{-0.31}$ & 1.34$^{+0.38}_{-0.35}$ \\
$C_{swift2}$  & 1.18$^{+0.33}_{-0.25}$  & 1.19$^{+0.30}_{-0.26}$\\
$C_{swift3}$  & 0.96$^{+0.31}_{-0.24}$ & 0.98$^{+0.29}_{-0.27}$ \\
$C_{swift4}$  & 0.96$^{+0.37}_{-0.27}$ & 0.96$^{+0.33}_{-0.25}$ \\
$C_{swift5}$ & 1.09$^{+0.31}_{-0.24}$  & 1.13$^{+0.30}_{-0.27}$ 
\enddata
\begin{tablenotes}{\footnotesize \textbf{Notes:} Refer to Tables \ref{tab:reflection} and \ref{tab:nhvar} for notes on parameter definitions and symbols.}
\end{tablenotes}
\end{deluxetable}
\vspace*{-15mm}

\begin{deluxetable}{c|c|c}

\tablecaption{Best fit parameters for the \nustar-only campaign of Mrk 477.}
\label{table:split_nustar}
\tablehead{\colhead{Model} & \colhead{\bor} & \colhead{\uxc} \\  }
\startdata
red $\chi^2$  & 0.94   & 0.94\\
$\chi^2$/d.o.f.  & 631.4/673  & 637.3/681 \\
$\Gamma$  & 1.57$^{+0.03}_{-0.06}$ & 1.80$^{+0.03}_{-0.05}$ \\
$N_{H,av}$  & 0.08$^{+0.02}_{-0.02}$ & $-$ \\
CF (Tor)  & 1.00$^{+u}_{-u}$  & $-$\\
CTKcover  & $-$ & 0* \\
$\sigma_{\rm tor}$ & $-$ &  27.8$^{56.2}_{-1.8}$ \\
Cos ($\theta_{Obs}$)  & 0.05$^{+u}_{-u}$ & 0.00$^{+u}_{-u}$\\ 
F$_s$ (10$^{-2}$)  & 6.53$^{+0.86}_{-0.15}$ & 16.1$^{+9.2}_{-1.8}$ \\
norm (10$^{-3}$) & 1.62$^{+0.19}_{-0.31}$ & 3.26$^{+0.11}_{-0.25}$ \\
\hline 
$N_{H,nus1}$ & 0.38$^{+0.05}_{-0.05}$ &  0.34$^{+0.05}_{-0.04}$  \\
$N_{H,nus2}$ & 0.48$^{+0.10}_{-0.02}$ &  0.46$^{+0.02}_{-0.03}$  \\
$N_{H,nus3}$ & 0.50$^{+0.10}_{-0.03}$ &  0.47$^{+0.03}_{-0.03}$  \\ 
$N_{H,nus4}$ & 0.48$^{+0.05}_{-0.02}$ &  0.43$^{+0.02}_{-0.02}$ \\
$N_{H,nus5}$ & 0.51$^{+0.08}_{-0.03}$ &  0.48$^{+0.03}_{-0.04}$  \\\hline
$C_{nus1}$ & 1* & 1*  \\
$C_{H,nus2}$ & 1.07$^{+0.03}_{-0.09}$ & 1.04$^{+0.10}_{-0.04}$  \\
$C_{H,nus3}$  & 1.04$^{+0.04}_{-0.10}$ &1.00$^{+0.10}_{-0.03}$  \\ 
$C_{H,nus4}$ & 1.20$^{+0.04}_{-0.15}$ & 1.13$^{+0.03}_{-0.10}$  \\
$C_{H,nus5}$ & 0.85$^{+0.09}_{-0.03}$ & 0.82$^{+0.31}_{-0.19}$ \\
\enddata
\begin{tablenotes}{\footnotesize \textbf{Notes:} Refer to Tables \ref{tab:reflection} and \ref{tab:nhvar} for notes on parameter definitions and symbols.}
\end{tablenotes}
\end{deluxetable}
\vspace*{-15mm}

\end{document}